\documentclass[journal]{IEEEtran}

\usepackage{nomencl}
\usepackage{amsmath}
\usepackage{amsfonts}
\usepackage{comment}
\usepackage[table]{xcolor}
\usepackage{tikz}
\usepackage{gensymb}
\usepackage{bm}
\usepackage{mathtools}
\usepackage[long]{optidef}
\usepackage{soul}
\usepackage{multirow}
\usepackage{booktabs,colortbl}
\usepackage[utf8]{inputenc}
\usepackage[T1]{fontenc}
\usepackage{balance}
\usepackage{hhline}
\usepackage{graphicx}
\usepackage{array}
\usepackage{subfigure}
\usepackage{color}
\usepackage{relsize}
\usepackage{cite}
\usepackage{comment}
\usepackage{epstopdf}
\usepackage{float}
\usepackage{stfloats}
\usepackage{tablefootnote}
\usepackage{tabularx}
\usepackage{threeparttable}
\usepackage{threeparttable} 
\usepackage{textcomp, gensymb}
\usepackage[linesnumbered,ruled,vlined]{algorithm2e}    
\usepackage{algorithmicx}   
\usepackage{algpseudocode}  
\SetKwComment{Comment}{$\triangleright$\ }{} 
\usepackage[en-US]{datetime2}
\usepackage{amssymb}

\usepackage{booktabs}
\usepackage{multirow}

\definecolor{o}{RGB}{0, 0, 0}
\definecolor{rev1}{RGB}{0, 0, 0}

\usepackage[pdftex,
pdfauthor={Linh Vu, Tuyen Vu, Thanh Long Vu, and Arurag Srivastava},
pdftitle={Multi-agent\ Deep\ Reinforcement\ Learning\ for\ Distributed\ Load\ Restoration},
pdfsubject={Load\ Restoration},
pdfkeywords={Deep reinforcement learning, invalid action masking, distribution systems, networked microgrids, multi-agent systems, load restoration, OpenDSS},
pdfproducer=Overleaf,
pdfcreator=pdfLaTeX,
pdfcreationdate={D:20220101120000},
pdfmoddate={\pdfcreationdate}]{hyperref}

\hypersetup{
	colorlinks,%
	citecolor=blue,%
	filecolor=black,%
	linkcolor=blue,%
	urlcolor=black
}

\graphicspath{{./figures/}}

\begin{document}

\title{Multi-agent Deep Reinforcement Learning for Distributed Load Restoration}

\author{
	Linh~Vu,~\IEEEmembership{Student Member,~IEEE,}
	Tuyen~Vu,~\IEEEmembership{Member,~IEEE,} \\
	Thanh~Long~Vu,~\IEEEmembership{Member,~IEEE,}
	and~Anurag~Srivastava,~\IEEEmembership{Fellow,~IEEE}
}

\markboth{}
{Vu \MakeLowercase{\textit{et al.}}: Multi-agent Deep Reinforcement Learning for Distributed Load Restoration}

\maketitle

\begin{abstract}
	This paper addresses the load restoration problem after power outage events. Our primary proposed methodology is using multi-agent deep reinforcement learning to optimize the load restoration process in distribution systems, modeled as networked microgrids, via determining the optimal operational sequence of circuit breakers (switches). An innovative invalid action masking technique is incorporated into the multi-agent method to handle both the physical constraints in the restoration process and the curse of dimensionality as the action space of operational decisions grows exponentially with the number of circuit breakers. The features of our proposed method include centralized training for multi-agents to overcome non-stationary environment problems, decentralized execution to ease the deployment, and zero constraint violations to prevent harmful actions. Our simulations are performed in OpenDSS and Python environments to demonstrate the effectiveness of the proposed approach using the IEEE 13, 123, and 8500-node distribution test feeders. The results show that the proposed algorithm can achieve a significantly better learning curve and stability than the conventional methods. 
\end{abstract}

\begin{IEEEkeywords}
	Deep reinforcement learning, invalid action masking, distribution systems, networked microgrids, multi-agent systems, load restoration, OpenDSS.
\end{IEEEkeywords}

\IEEEpeerreviewmaketitle

\section*{Nomenclature}
\addcontentsline{toc}{section}{Nomenclature}
\begin{IEEEdescription}
	[\IEEEusemathlabelsep\IEEEsetlabelwidth{$V_1,V_2,V_3$}]
	\item[$P_{ij}$] Rated power of load $j$ (downstream of switch $i$)
	\item[$w_{ij}$] Weighting factor of load $j$ (downstream of switch $i$)
	\item[$n_i$] Binary variable, represents the status of switch $i$ (0 means switch $i$ is open and 1 means switch $i$ is close)
	\item[$V_u$] Voltage at bus $u$
	\item[$p_u$] Active power injected at bus $u$
	\item[$q_u$] Reactive power injected at bus $u$
	\item[$s_u$] Apparent power injected at bus $u$
	\item[$s_{mn}$] Apparent power from bus $m$ to bus $n$
	\item[$o$] Current observation of the system
	\item[$o'$] Next observation of the system
	\item[$r$] Current reward
	\item[$a_i$] Current action of agent $i$
	\item[$A_i$] Action space of agent $i$
	\item[$\textbf{a}$] Joint action of all agents
	\item[$m$] Invalid action masking information
	\item[$\gamma$] Discounted value
	\item[$\alpha$] Learning rate
	\item[$\epsilon$] Epsilon value
	\item[$\lambda$] Decay value
	\item[$Q(s,a)$] Q value of ANN model
	\item[$\theta$] Weights and biases of ANN model.
\end{IEEEdescription}

\newpage 
\section{Introduction}

\IEEEPARstart{I}{n} recent years, weather patterns have varied unpredictably. Unfortunately, the power systems are not adequately prepared to face extreme weather events. For example, the Texas freeze in February 2021 impacted 4.5 million power users and resulted in \$130 billion in economic damages \cite{BUSBY2021102106}; Hurricane Isaias in August 2020 affected more than 6.4 million customers; the California 2019 Power Shutoffs in October 2019 impacted 1.6 million customers and the New England Storm in October 2019 affected 1.25 million customers. \textcolor{rev1}{In certain worst-case scenarios, it can take an extended period of time for power to be restored through the utility grid and made available to customers.}

A microgrid is one promising solution that can produce immediate power to specific customers in the event of a utility power outage. After a fault, microgrids can switch from grid-connected mode to islanded mode and operate independently. However, since the power from microgrids is limited, they cannot provide a full power supply for all customers, so critical loads must be identified and prioritized. In order to transfer power only to critical loads, it is vital to determine the optimal sequence to regulate the circuit breakers. Solving the optimal pattern of circuit breaker usage is referred to as the load restoration problem. Load restoration is a complex problem as its properties involve multi-objective, non-linear, and combinatorial selections. To tackle the load restoration problem, several approaches have been investigated. Recent studies on load restoration may be divided into two categories: optimization-based and machine learning-based methods.

\textbf{Optimization-based approach:} The Spanning Tree Search approach is presented in \cite{6781027}, in which the objective function is formulated to minimize the number of switching actions and maximize the restored power. Although this formulation has reduced the operating stress on the switches, it failed to address the essential aspect of load priority. The authors in \cite{7447808} have overcome this limitation by reformulating the problem into two objective functions to integrate the weighting of loads and minimize the voltage variation of the power network. Furthermore, a multi-stage restoration is proposed in \cite{7428966}, where load priority, number of switching actions, and voltage variation are combined into the objective function. In order to formulate the problem where some switching variables are represented as integers and others as continuous variables, further optimization-based approaches employing mixed-integer linear programming (MILP) methods are described in \cite{8409997, 9488172, 8879675, 8283770} or mixed-integer non-linear programming (MINLP) methods in \cite{7924370, 9094201, 4162603}. The results from these papers using mixed-integer methods have indicated that a globally optimal solution can be achieved, but the computation time presents a significant challenge, especially for real-time decision-making. In \cite{8845587}, the authors presented a novel formulation that has reduced variables, constraints, and computation time. Other aspects of load restoration problems have also been investigated in \cite{8587147, 8745959, 8731667} that include repair time, renewable energy availability, and the optimal path for sending crews to the fault fields.

\textbf{Learning-based approach:}
\textcolor{rev1}{The optimization-based methods are primarily restricted in deterministic environments or bounded uncertainty conditions, whereas often there are uncertainties in the data and model that are not bounded and must be taken into account, such as load randomness, cold load pickup, imperfect renewable forecast, and mobile energy resources \cite{7779481, 9576535, 9903581, 9282132}. Moreover, optimization-based methods mainly produce the final configuration topology, largely omitting the sequence needed to regulate the switches \cite{6305493, https://doi.org/10.1049/iet-gtd.2018.5624}. Furthermore, optimization-based approaches have an additional disadvantage due to the non-convex nature of power flow constraints and the requirement for a complete and accurate mathematical model of the environment. As demonstrated in \cite{JIANG2017127}, the computing complexity of classical MILP formulations can make them infeasible to deploy. Since the learning-based approach uses trained models to make decisions, the computational time can be significantly reduced because it only involves feed-forward calculations to make decisions. To take advantage of learning-based approach and to overcome such limitations of optimization-based approach, reinforcement learning (RL) has recently been gaining attention. The authors in \textcolor{rev1}{\cite{5618962, 8862818, 5871714, 7173055} }introduced Q-learning, an off-policy, model-free reinforcement learning algorithm. The idea of Q-learning is to create a finite table of states and actions, then seek the best action given the current system states. Although this method can ensure policy convergence and an optimal solution \cite{melo2001convergence}, it is very burdensome to use at a large scale on complex power systems since the number of states and actions will grow exponentially as the number of switches increases.} \textcolor{rev1}{To address this drawback, a Deep Q-learning approach is proposed in \cite{9940073}. Instead of forming a table to store state-action pairings, Deep Q-learning utilizes an artificial neural network (ANN) as a functional approximation of state-action pairs and trains the Deep Q-networks to generate the optimal decision. ANNs are also used as one of their algorithms' components in \cite{9705112, 9282132}, where two ANNs known as the critic network and actor network are implemented in the Deep Deterministic Policy Gradient (DDPG) and Twin Delayed Deep Deterministic Policy Gradient (TD3) algorithms.} Those algorithm uses a critic network to estimate the action-value function and an actor network to learn the optimal policy of the load restoration agent. To deal with the uncertainty of available power from renewable energy during the restoration process, the authors in \cite{9302946} developed a reinforcement learning algorithm based on evolution strategies. Furthermore, work in \cite{9345996} integrated the asynchronous information of available generator power resources into the DQN agent in the training phase.

\textbf{Multi-agent learning:} Although the above-mentioned reinforcement learning algorithms have been carried out to solve the load restoration problem, they are performed in a centralized manner, where only one centralized agent interacts with the environment at each time step. However, such a centralized machine learning-based control paradigm requires communication links from the central controller to all remote locations. This can potentially incur communication failures or cyberattacks on the controller and communication links. Also, participants need to upload the necessary information to the controller. Thus some private information may be at risk of leaks. Most critically, training the centralized RL on large-scale systems with thousands of nodes would require a prohibitively long training time to train an excessive number of neural network parameters. Finally, many problems in real-world environments require more than one agent to be fully solved; examples include unmanned aerial vehicles \cite{9679711},  traffic flow \cite{8667868}, and more \cite{9189872,8653482,9536423,9157356,jpm12020309}.  In \cite{10.5555/3091529.3091572}, the author shows that by using a multi-agent reinforcement learning (MARL) platform, the hunters (agents) can collaborate to capture prey faster than a single agent by sharing their learned knowledge. According to \cite{pmlr-v97-iqbal19a}, authors emphasize that MARL can accelerate the learning process since the agents not only learn from the environment but also from each other. \textcolor{rev1}{In \cite{7088641}, the multi-agent concept was firstly leveraged to restore service in the distribution systems by treating different types of electrical devices (such as load, aggregator, distributed generator, and switch) as individual agents. These agents could communicate with each other to implement a heuristic rule-based algorithm. However, this approach focused primarily on the communication aspect between agents and did not utilize learning-based techniques. In contrast, the work presented in \cite{9427147} utilized deep reinforcement learning to enhance power system resilience by using the shunt reactive power compensators as control objectives. To control the status of circuit breakers, another approach used in \cite{9508140} was a graph-based method, where each agent must depart from a depot and travel sequentially along the energization path based on features from graphs. However, this paper's limitation was that the agents had limited autonomy in deciding their actions.}

In this paper, we adopted the advantages of a deep MARL platform to restore critical loads in distribution systems that can be modeled as networked microgrids. In a mixed ownership environment, each microgrid can belong to either a utility or non-utility owner, and hence, no centralized controller is authorized to control all microgrids. In this context, it is natural to adopt the multi-agent framework, in which each microgrid has one reinforcement learning agent. \textcolor{rev1}{These RL agents in networked microgrids can aggregate their information in a centralized training setting to learn optimal control policies. The collaboration of multi-agents consists of: (1) sharing actions taken by other agents, (2) sharing invalid actions to avoid, and (3) sharing the reward they contribute to the global reward function.}
\\\textbf{The following are the paper's main contributions:}

\begin{itemize}
	\item To the best of our knowledge, this is the first time a multi-agent deep reinforcement learning algorithm has been developed and adopted to solve the load restoration problem.
	\item The paper proposes a novel enhancement into the MARL of an invalid action masking technique to improve the learning performance and convergence against physical constraint violations to ensure safe load restoration.
	\item The paper compares the convergence curves and optimal results between single-agent and multi-agent RL and also provides a comprehensive analysis of the importance of the invalid action masking technique.
\end{itemize}

The remainder of this paper is organized as follows. Section \ref{SectionII} is divided into four subsections. We formulated the load restoration problem as an optimization in \ref{Load Restoration Problem}. Then a multi-agent DQN framework is introduced to control multiple microgrids inside a complex power system in \ref{Multi-agent Formulation}. \ref{Epsilongreedy method} describes the epsilon-greedy method to balance exploration and exploitation. Lastly, \ref{invalid action masking} presents the invalid action masking technique and the main algorithm in detail. Section \ref{SectionIII} shows the experimental results using IEEE standard test feeders.

\section{Multi-agent Reinforcement Learning for Load Restoration}
\label{SectionII}
\begin{figure}[!t]
	\centering
	\includegraphics[width=1.0\linewidth]{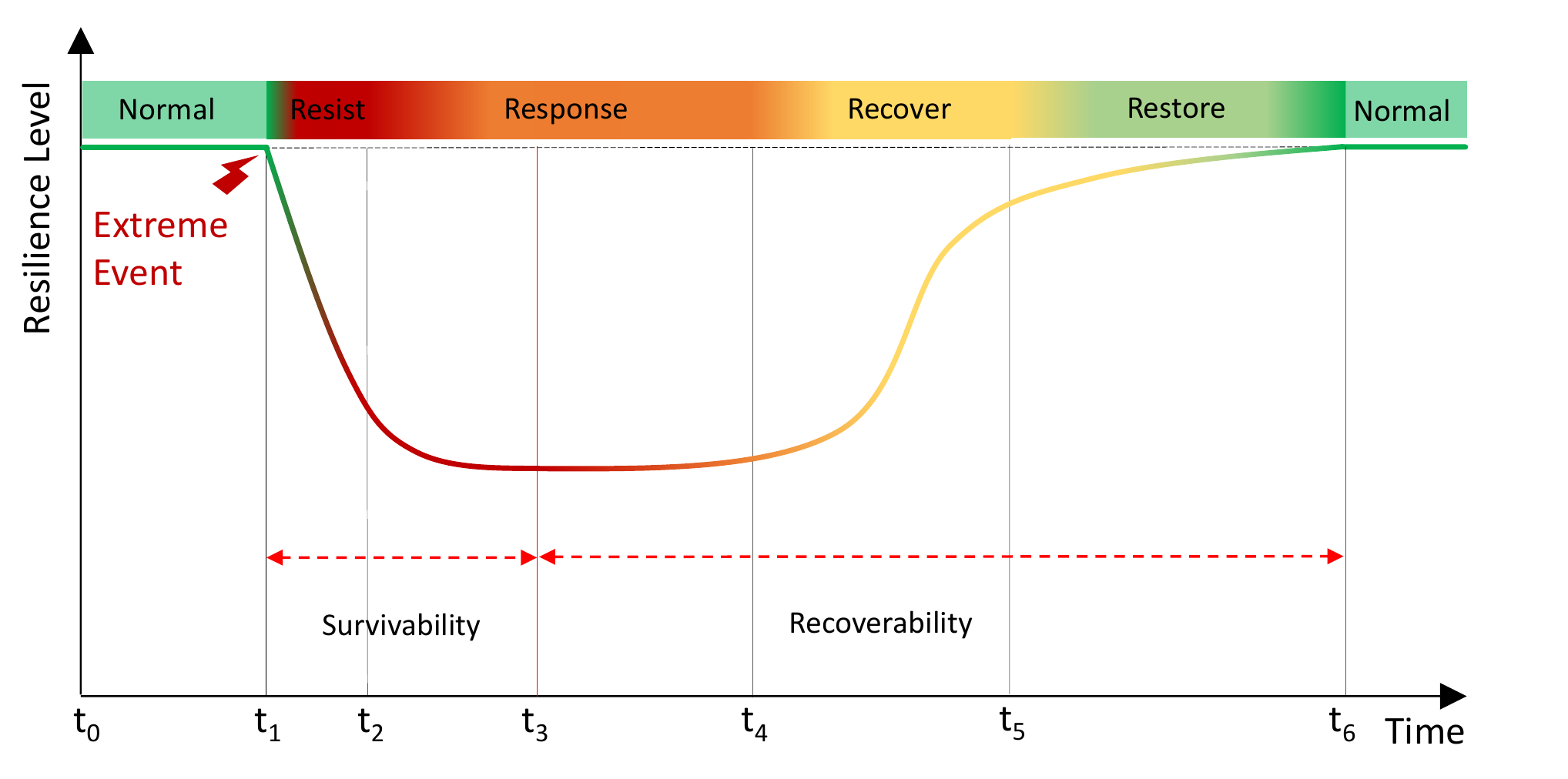}
	\caption{The concept of resiliency curve during an extreme event.\cite{9205672}}
	\label{fig:200}
\end{figure}

\noindent The power system's capability to resist and minimize the magnitude or duration of extreme events, including the ability to predict, absorb, adapt to, or quickly recover from such catastrophes, is called resilience\cite{9121348}. Several publications \cite{en14030694,en11092272,JUFRI20191049,7893706,7091066} defined the resilience curve to evaluate the performance of the power system during a disaster. Illustrated in Fig. \ref{fig:200}, the resilience level is used to measure the ability of a power system under an extreme event. This process can be divided into several stages. In this paper, our proposed algorithm aims to focus on the \textbf{recoverability} stage represented as $t_4\ \text{to}\ t_5$, where the proposed algorithm will work on maximizing the critical loads within the available power limitations.

\subsection{Load Restoration Problem}
\label{Load Restoration Problem}
\noindent During the restoration period, power channels from the utility grid to consumers are disrupted, requiring microgrid systems as alternate sources to meet the power demand of critical loads. These power paths are defined by the status of circuit breakers, which were installed in the power systems to manage the power supply to downstream loads.

The primary objective function:
\begin{maxi!}|l|[3]
{n_i}{\sum n_i \sum P_{ij}w_{ij}}{}{{\labelOP{eq:2010}}}
{}{}
\addConstraint{\sum P_{loss}+\sum n_i \sum P_{ij}}{\leq \sum P_{gen}}{\labelOP{eq:2011}}
\addConstraint{V^{min}}{\leq V_u\leq V^{max}\labelOP{eq:2012}}
\addConstraint{p_{min}}{\leq p_u\leq p_{max}\labelOP{eq:2013}}
\addConstraint{q_{min}}{\leq q_u\leq q_{max}\labelOP{eq:2014}}
\addConstraint{p^2_{mn}+q^2_{mn}}{\leq s^2_{mn.}}{\labelOP{eq:2015}}
\end{maxi!}

The main objective function is formulated in the form of optimization-based equations (\ref{eq:2010}), where the purpose is to find the maximum value of the restored load. Several safe constraints of the physical power system are satisfied by (\ref{eq:2011} - \ref{eq:2015}) during the restoration process. In (\ref{eq:2010}), $P_{ij}$ and $w_{ij}$ are rated power and weighting factor (that represents the load's priority) of the downstream load $j$ of the circuit breaker $i$, respectively. Binary variable $n_i$ represents the status of circuit breakers $i$ ($0$ means circuit breaker is open, and vice versa). Constraint (\ref{eq:2011}) ensures the power balance between generators and loads. Constraint (\ref{eq:2012}) guarantees the operating voltage at each bus to be within a safe range of the nominal $V_i$ value. \textcolor{rev1}{Constraint (\ref{eq:2013} - \ref{eq:2014}) represents the active power ($p_u$) and reactive power ($q_u$) output limits for each local generator. Finally, equality constraint (\ref{eq:2015}) represents the sending-end complex power from bus $m$ to bus $n$.}

\subsection{Multi-agent DQN Formulation for Load Restoration}
\label{Multi-agent Formulation}
\noindent Based on the problem formulation in \ref{Load Restoration Problem}, this paper proposes a multi-agent deep reinforcement learning framework for load restoration. We consider that a distribution power system can be divided into multiple microgrids. The division depends on the total rated power of loads, the number of switches, and the available power of generators. In this framework, each agent is in charge of a distinctive microgrid and is responsible for the status of all switches inside that microgrid.

\begin{figure}[!t]
	\centering
	\includegraphics[width=1.0\linewidth]{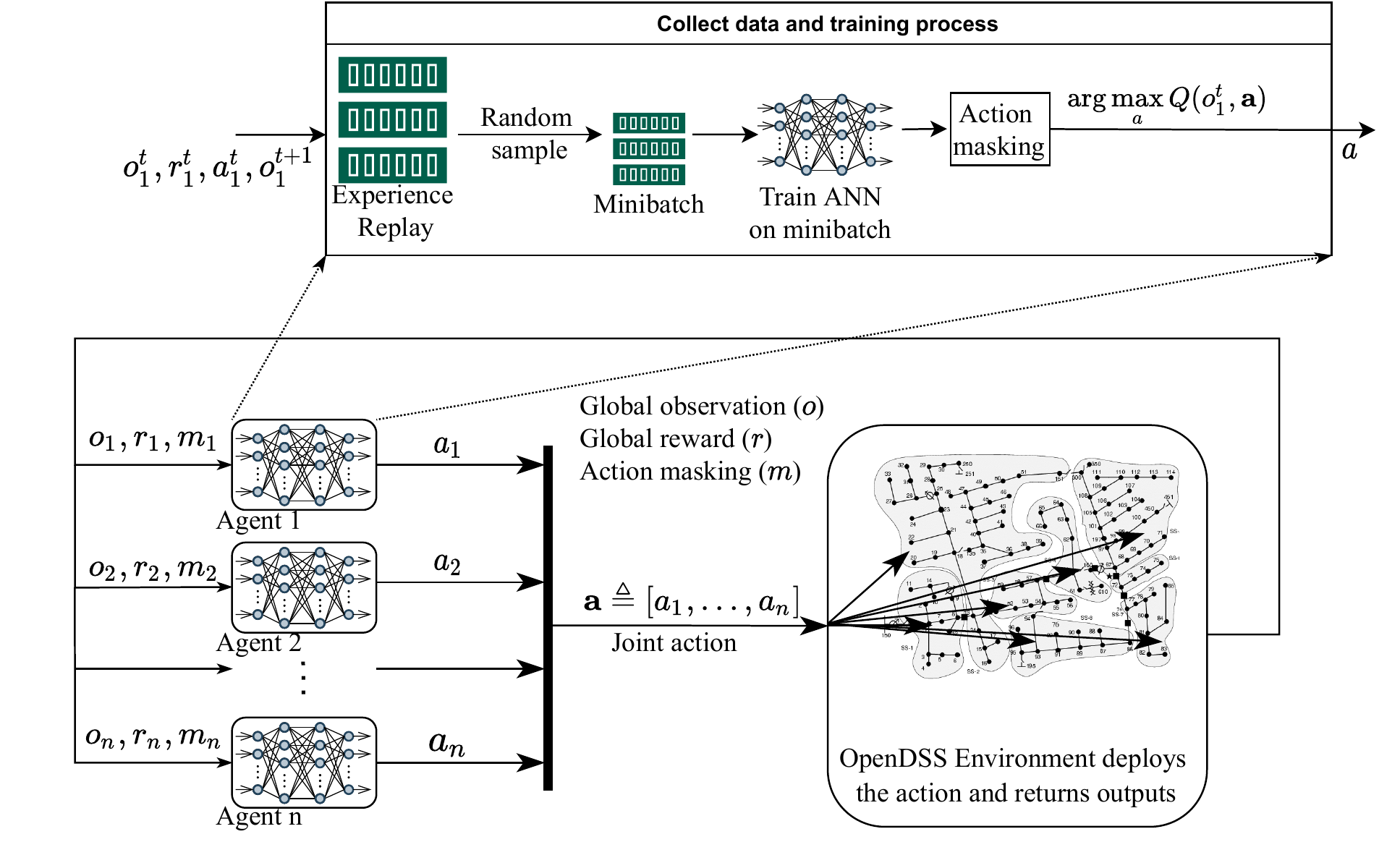}
	\caption{The overall diagram of multi-agent DRL framework.}
	\label{fig:201}
\end{figure}

This multi-agent algorithm is an enhancement based on single-agent deep reinforcement learning algorithms \cite{9345996}. We formulate the load restoration problem as a Markov decision process (MDP) \cite{LITTMAN20019240}. Fig. \ref{fig:201} illustrates the overall diagram of the multi-agent framework. The proposed algorithm uses $n$ agents ($n\geq2$), where each agent is an artificial neural network model, to learn from the environment and make decisions. In the \textcolor{rev1}{centralized} training phase, each agent chooses their own action simultaneously (i.e., agent 1 sends action $a_1$, agent 2 sends action $a_2$, etc.) at the beginning of every time-step from their action spaces. \textcolor{rev1}{The epsilon-greedy policy (\ref{Epsilongreedy method}) is utilized for decision-making by each agent, which allows for either random selection or following their current policies $ a_{i,t} \sim \pi (\cdot | s_{i,t}) $. These individual actions are then aggregated into a joint action denoted as $\textbf{a} \triangleq [a_1,\dots,a_n]$, as described in \cite{Buoniu2010MultiagentRL}. The environment receives the joint action and deploys the corresponding actions to each microgrid, resulting in a transition of the system's state to the next state $s_{t+1} \sim \rho(s_{t+1} | s_t, \textbf{a}_t)$. Each agent then receives an immediate scalar reward, and the resulting system state (observation ($o$), reward ($r$), action masking ($m$)) is provided as feedback to the agents.}

After each training time-step, each agent stores a tuple information of observation, reward, action, and next observation ($o^t, r^t, a^t, o^{t+1}$) to accumulate the knowledge. The agents repeat this process multiple times until they have sufficient information to tackle the problem. Below are \textcolor{rev1}{detailed} descriptions of the environment, action space, observation, and reward function for problem formulation.

\textbf{Environment}: In this framework, the environment is the distribution system, which is divided into multiple microgrids as shown in Fig. \ref{fig:201}. Controllable circuit breakers are installed for test feeders to maneuver the power supply paths to the loads. Each time a circuit break is activated, the states of the system will be changed accordingly. Therefore, the power flow problem needs to be solved to determine the states of the system following the deployment of an action and to guarantee that all constraints will be met. Any power system simulation tools such as OpenDSS \cite{dugan2018open}, MATLAB-Simulink, or MATPOWER \cite{5491276} can be used to simulate the environment.

\textbf{Action space}: \textcolor{rev1}{The action at each time-step of an agent refers to turning on or off one circuit breaker inside its microgrid. In this paper, we assume that circuit breakers can be monitored and operated ideally following the commands from the controller.} Therefore, the control action of agent 1 can be represented as an array of binary values $a_1 = [cb1_{on}\ cb1_{off}\ cb2_{on}\ cb2_{off} \dots cbn_{on}\ cbn_{off}]$, where $n$ is the number of circuit breakers in microgrid 1. For instance, agent 1 selects to turn on circuit breaker 2, then $a_1 = [0\ 0\ 1\ 0\ \dots\ 0\ 0]$. Note that each agent will have a different size of action space depending on the number of circuit breakers (switches) inside its microgrid. The joint action is a combination of actions sending from all agents \textcolor{rev1}{($\textit{\textbf{a}} \triangleq [a_1,\dots,a_n]$) to the environment.}

\textbf{Observation}: \textcolor{rev1}{In this framework, each agent is able to access its local observation at every time step as shown in Fig. \ref{fig:201}. This observation consists of the status of all circuit breakers within its zone and can be represented as an array of binary values, denoted as $\textit{\textbf{o}} = [o_1\ o_2\ \dots\ o_m]$, where $m$ represents the number of circuit breakers under the agent's responsibility. Note that when $o_1 = 1$, it indicates that circuit breaker 1 is in a closed state, and conversely, when $o_1 = 0$, it means that circuit breaker 1 is in an open state.}

\begin{figure}[!t]
	\centering
	\includegraphics[width=0.9\linewidth]{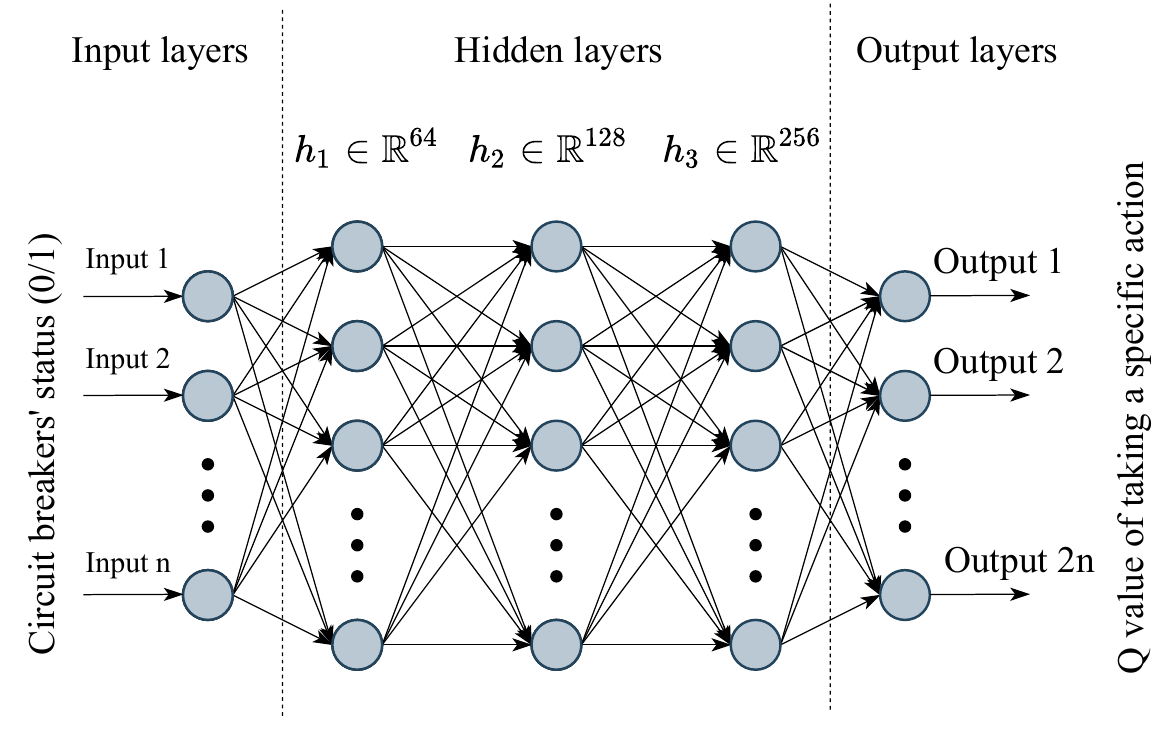}
	\caption{\textcolor{rev1}{An example of an agent's ANN structure within a multi-agent framework.}}
	\label{fig:202}
\end{figure}

\textbf{Reward function}: In this framework, all the agents will share the same global reward function, which is shown in (\ref{eq:2020}). Similar to the main objective function in \ref{eq:2010}, this reward function is designed to encourage the agents trying to maximize the restored power of loads. It should be noted that we normalized the reward function (re-scale to between 0 and 1) in order to stabilize the training.
\begin{align}
	r = \frac{\sum n_i \sum P_{ij}w_{ij}}{\sum P_{ij}}.
	\label{eq:2020}
\end{align}
\indent Fig. \ref{fig:202} goes into detail about \textcolor{rev1}{the Deep Q-network architecture of an agent within a multi-agent framework. The input of ANN is the current observation of this agent, which is the status of all the circuit breakers within its microgrid}, while the output is Q-values for selecting the available actions. The output of ANN has its number depends on the number of circuit breakers that the agent manages and represents the benefit of selecting the actions. The hidden layers are designed depending on the complexity of the problem to store the learned policies.

\textcolor{rev1}{This paper incorporates the experience replay technique to break harmful correlations and enable agents to learn more effectively by recalling rare occurrences. Specifically, each agent has its own experience replay data during training, consisting of multiple tuples (observation, reward, action, next observation) learned during exploration and stored for each agent. To utilize this data for training, the training process randomly samples small mini-batches from the larger experience replay dataset, as shown in Fig. \ref{fig:201}.}

\begin{figure}[!t]
	\centering
	\includegraphics[width=1.0\linewidth]{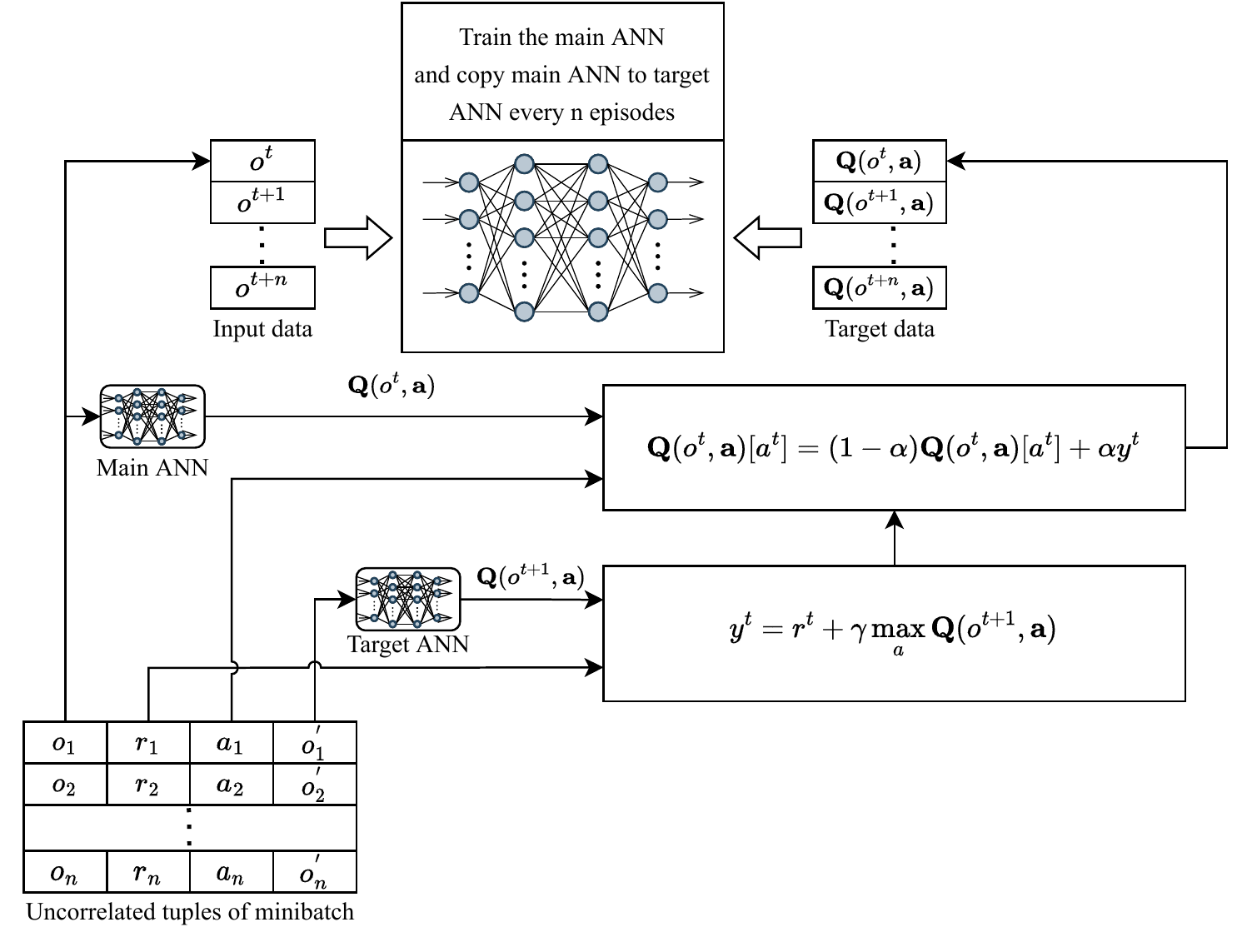}
	\caption{\textcolor{rev1}{The training process of an agent within a multi-agent framework.}}
	\label{fig:203}
\end{figure}

Fig. \ref{fig:203} shows the overall diagram of the training process. The RL agents collect data from the trial and error process until they perform sufficient numbers of actions in the OpenDSS environment. For the training process, two ANNs are used for each agent, namely the main ANN and target ANN. They have the same initialization and the weights of the main ANN will be updated periodically into the target ANN. According to \cite{mnih_human-level_2015, https://doi.org/10.48550/arxiv.2106.02613}, the implementation of two ANNs for agents helps slow down learning, strengthens the algorithm, and improves the effectiveness of agent environment exploration.

Based on the current reward and the next observation, we can calculate the maximum value in the future that the agent can take after choosing the current action at time step $t$ using (\ref{eq:2030}). Value $\gamma$ is called the discounted value, and it is used to decide the importance of rewards in the distant future relative to those in the immediate future.
\begin{align}
	y^t = r^t + \gamma \max_a Q(o^{t+1},\textbf{a})
	\label{eq:2030}
\end{align}
\indent Equation (\ref{eq:2060}) is used to update the $Q_{predicted}(s,a)$ value during the training to track the $Q_{target}(s,a)$ value gradually. Value $\alpha$ is known as learning rate value to decide the changing rate of $Q_{target}(s,a)$ compared to $Q(s,a)_{predicted}$.
\begin{align}
	Q_{target}(s,a) = (1 - \alpha)Q(s,a)_{predicted} + \alpha y^t
	\label{eq:2060}
\end{align}
\indent To train the model, a loss function is defined as shown in (\ref{eq:2040}) to perform gradient descent steps.
\begin{align}
	L(\theta) = \mathbb{E}[(Q_{target}(s,a) - Q_{predicted}(s,a))^2]
	\label{eq:2040}
\end{align}
\indent This equation represents the squared difference between the $Q_{target}(s,a)$ and the $Q_{predicted}(s,a)$.

\indent For each individual tuple of the mini-batch, we can calculate a pair of input data as shown in Fig. \ref{fig:203}, which is the state of the system and output data. Based on the input and target data, the model can be trained to fit the data as the Q-value will be updated during the exploration process.

\indent \textcolor{rev1}{In the decentralized execution phase, the agents utilize the trained models to make decisions based on their local observations without the invalid mask technique. This approach offers the added benefit of enhancing data privacy and security since the data is processed locally on each networked-microgrid and is not transmitted to a central controller. This is particularly crucial in scenarios where sensitive data is at risk of being compromised by cyberattacks.}

\subsection{Epsilon-greedy Method}
\label{Epsilongreedy method}
\noindent \textcolor{rev1}{The epsilon-greedy approach \cite{https://doi.org/10.48550/arxiv.1312.5602} is selected to balance exploration and exploitation over the learning course. This method is motivated by the fact that during the initial training phase, agents lack information about the environment and must prioritize exploration over exploitation. Once the agents have gained sufficient knowledge through training the ANN models, they can rely on that knowledge to inform their decision-making.}

\textcolor{rev1}{At the beginning of the algorithm, the epsilon value is initialized to 1. To gradually decrease the epsilon value $\epsilon$ to a minimum value of $\epsilon_{min}$ over episodes, (\ref{eq:2070}) is utilized. The decay value ($\lambda$) determines the rate of reduction, while $\epsilon_{max}$ and $\epsilon_{min}$ define the upper and lower limits of epsilon, respectively.}
\begin{align}
	\epsilon = \epsilon_{min} + (\epsilon_{max} - \epsilon_{min}) e^{-\lambda \text{episode}}
	\label{eq:2070}
\end{align}
\subsection{Invalid Action Masking Technique}
\label{invalid action masking}
\noindent One of the biggest challenges in solving the load restoration problem is that we both need to maximize the main objective function as well as ensure all power flow constraints (\ref{eq:2013} - \ref{eq:2015}) are satisfied at all times.

The traditional approach is to put a large negative reward into the reward function each time the agents make a decision that would violate any constraint as described in \cite{10.5555/3312046,YOO2021487}.

The algorithm can perform well with this approach for a small distribution system with a modest number of switches. In this case, the reward function can be written as in (\ref{eq:2080}).
\begin{align}
	r = \begin{cases}
		    \sum n_i \sum P_{ij} w_{ij} & \text{All constraints satisfy} \\
		    M                           & \text{Any constraint violates}
	    \end{cases}
	\label{eq:2080}
\end{align}
\indent With M being a large negative value, agents are warned not to choose undesirable actions. One drawback of this approach is that M will vary depending on the particular problem. In most cases, M needs to be well-designed based on experience. This approach becomes even more complex when the system at \textcolor{rev1}{some certain states has} a very small number of actions that can satisfy all the constraints, cost the agent  a huge amount of time to learn, or even fail to learn \cite{https://doi.org/10.48550/arxiv.1910.12134}. \textcolor{rev1}{As demonstrated in \cite{https://doi.org/10.48550/arxiv.1912.06680}, DOTA2 is an example of an environment where agents are presented with up to 1,837,080 actions to choose from their action spaces at each time-step. However, most of these actions are irrelevant to the task at hand, resulting in a significant amount of wasted time if agents attempt to learn them all \cite{9619076}. Therefore, it's crucial to identify which actions should be ignored in order to optimize the learning process.}

To deal with this type of problem, AlphaStar and OpenAI 5 proposed an invalid action masking technique in their research framework \cite{hoffman2020acme, Vinyals2019GrandmasterLI}, which is a trick to mask out meaningless actions in the full action space of each agent.

\begin{figure}[!t]
	\centering
	\includegraphics[width=0.9\linewidth]{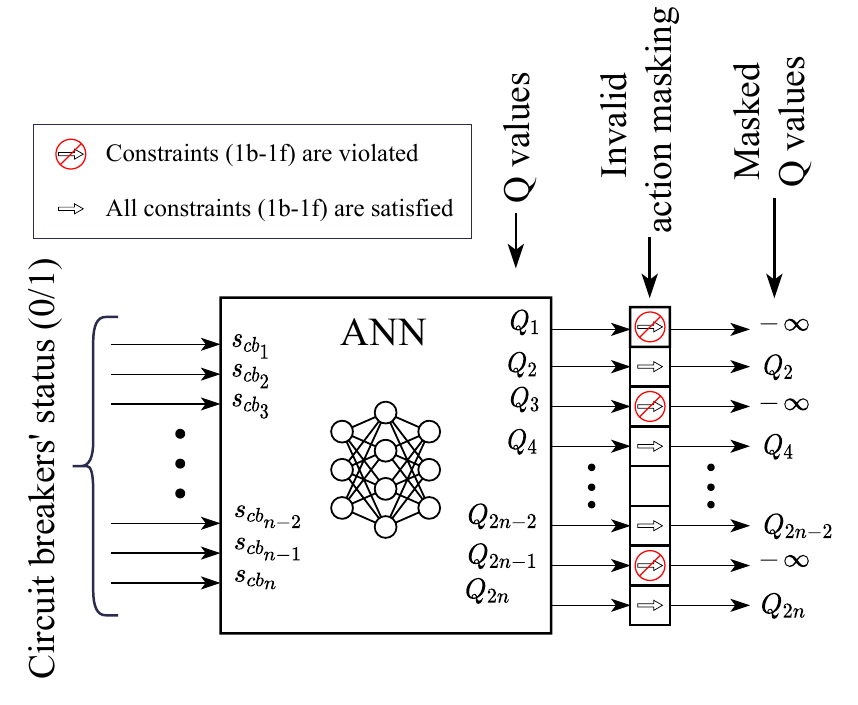}
	\caption{Visualization of action masking.}
	\label{fig:205}
\end{figure}

Fig. \ref{fig:205} shows the mechanism of the invalid action masking. With a given state as the input to each agent's ANN, the ANN will output the Q-value of picking actions corresponding to each circuit breaker turning on or off. The deep Q-learning algorithm uses the $argmax$ function to choose the action that raises the highest reward value across the available actions. The basic principle behind action masking is that it will obstruct this process by extracting invalid action information from the environment at each time-step and creating a mask that will inform the agents which one they should not take at that time. In \cite{Huang_2022}, the authors presented the theory of action masking and how to implement the action masking technique with the \textcolor{rev1}{Proximal Policy Optimization} (PPO) algorithm, which is a gradient-based policy algorithm by renormalizing the distribution of output Q-value after removing the action that would violate the constraints. In this study, we implemented invalid action masking by using the deep Q-learning method, which is a value-based policy technique, by assigning a big negative value ($-\infty$) to actions that violate the environment and then reselecting the action with the highest value after removing the first highest one.

Because our challenge involves a multi-agent system, a joint action should be considered to send to the environment in order for each agent to know if their actions would break the constraints or not. In case the joint action violates the environment constraints, a random agent will be selected to remove its highest action value and reselect the action from the agents again. In the next action selection, this would help reduce the restored power as well as the probability of violating the constraints. This process will be repeated until the agents choose a joint action that will not violate any constraint.

The above sections have introduced the essential keys of the algorithm. The detailed version of Algorithm~\ref{algo:ddrl} describes distributed DQL approach for load restoration.

\begin{algorithm}
	Initialize total number of agents $m$\;
	Initialize $\epsilon_0,\ \epsilon_{max},\ \epsilon_{min},\ \lambda$\;
	Initialize experience replay size and mini-batch size\;
	Initialize training episode $n_0,\ n_{max}$\;
	Initialize total training reward $R$\;
	Initialize step to update models $k_0,\ k_{max}$\;
	Initialize step within an episode $u_o,\ u_{max}$\;
	\Repeat{\emph{m agents are initialized}}
	{
		Initialize the main model for agent $i$\;
		Initialize the target model for agent $i$\;
		Copy weights from the main model agent $i$ to the target model agent $i$\;
		Initialize replay experience tuple for agent $i$\;
	}
	\For{$n_0 = 0$ \text{to} $n_{max}$}
	{
		$R = 0$; \hfill \Comment{Reset episode reward}
		Reset the OpenDSS Environment\;
		\While{$u_0 \leq u_{max}$}
		{
			$u_{0}\mathrel{+}=1$;\ $k_{0}\mathrel{+}=1$\;
			Randomly generate $\epsilon_i \in (0,1)$\;
			\If{$\epsilon_i \leq \epsilon_0$} 
			{
				\Repeat{\emph{$m$ agents decide their actions}}
				{
					Randomly select action $a_i$ for agent $i$ ($a_i \in A_i$)\; \label{lookback00}
					Joint actions from agents\;
				}
				\While{Invalid action masking == False}
				{
					Jump to step (\ref{lookback00})
				}
			}
			\Else
			{
				\Repeat{\emph{$m$ agents decide their actions}}
				{
					Select action $a_i$ for agent $i$ ($a_i \in A_i)$ based on their current policies\; \label{lookback02}
					Joint actions from agents\;
				}
				\While{Invalid action masking == False}
				{
					Randomly select an agent $j\ (j \in [1,m])$\;
					Assign the maximum Q-value of agent $j$'s current output to $-\infty$\;
					Jump to step (\ref{lookback02}) with modified agent $j$'s output\;
				}
			}
			OpenDSS deploys joint action then returns new observation ($o'$) and immediate reward ($r$)\;
			$R\mathrel{+}=r$\;
			\Repeat{\emph{$m$ agents are appended}}
			{
				Append $(o, a_i, r, o')$ into experience replay tuple of agent $i$\;
			}
			$o \gets o'$\;
			Train main models using experience replay (\ref{eq:2030}$-$\ref{eq:2060})\;
			\If{$(k_{o} \bmod k_{max}) == 0$}
			{
				Copy weights from main model to target model of each agent\;
			}
		}
		Reduce $\epsilon_0$ using (\ref{eq:2070});
	}
	\caption{Distributed deep reinforcement learning algorithm for load restoration}
	\label{algo:ddrl}
\end{algorithm}

\section{Case Studies}
\label{SectionIII}
\noindent The algorithm is verified by using an Intel(R) Core(TM) i7-10700 CPU (2.90 GHz processor and 16.0 Gb RAM). \textcolor{rev1}{To validate the scalability of the algorithm, we investigate three systems of varying sizes: small (IEEE 13-node), medium (IEEE 123-node), and large (IEEE 8500-node) systems.} The algorithm is coded in a Python environment and communicated with OpenDSS software using the  "py-dss-interface" package \cite{radatzvianalondero}.

\subsection{IEEE 13 Nodes}

\begin{figure}[!t]
	\centering
	\includegraphics[width=0.9\linewidth]{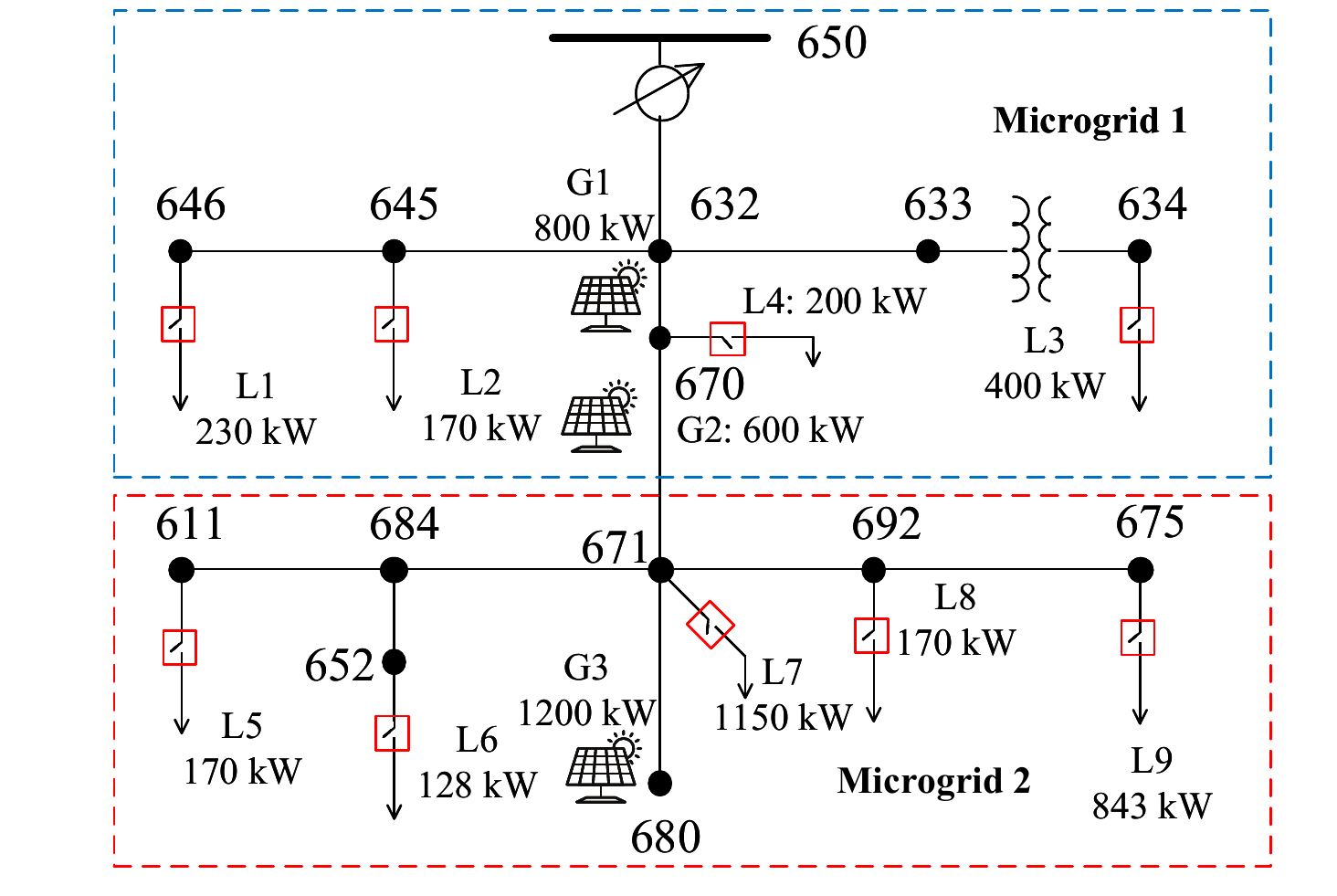}
	\caption{13-node networked microgrids configuration.}
	\label{fig:300}
\end{figure}

\noindent In Fig. \ref{fig:300}, the distribution system is divided into two microgrids corresponding to two agents. Microgrid 1 has four loads $L_{MG1} = [230,170,400,200]$ kW and microgrid 2 has five loads $L_{MG2} = [170,128,1150,170,843]$ kW. During normal operation, the system needs to provide 3461 kW to serve the full power of loads. Three solar power sources with a total capacity of 2600 kW are installed on the distribution system. After a major power outage, we assume that all circuit breakers are in the open state, and the algorithm needs to find a sequence of turning them on and off to maximize the restored power while still satisfying physical constraints.

\begin{figure}[!t]
	\centering
	\includegraphics[width=0.9\linewidth]{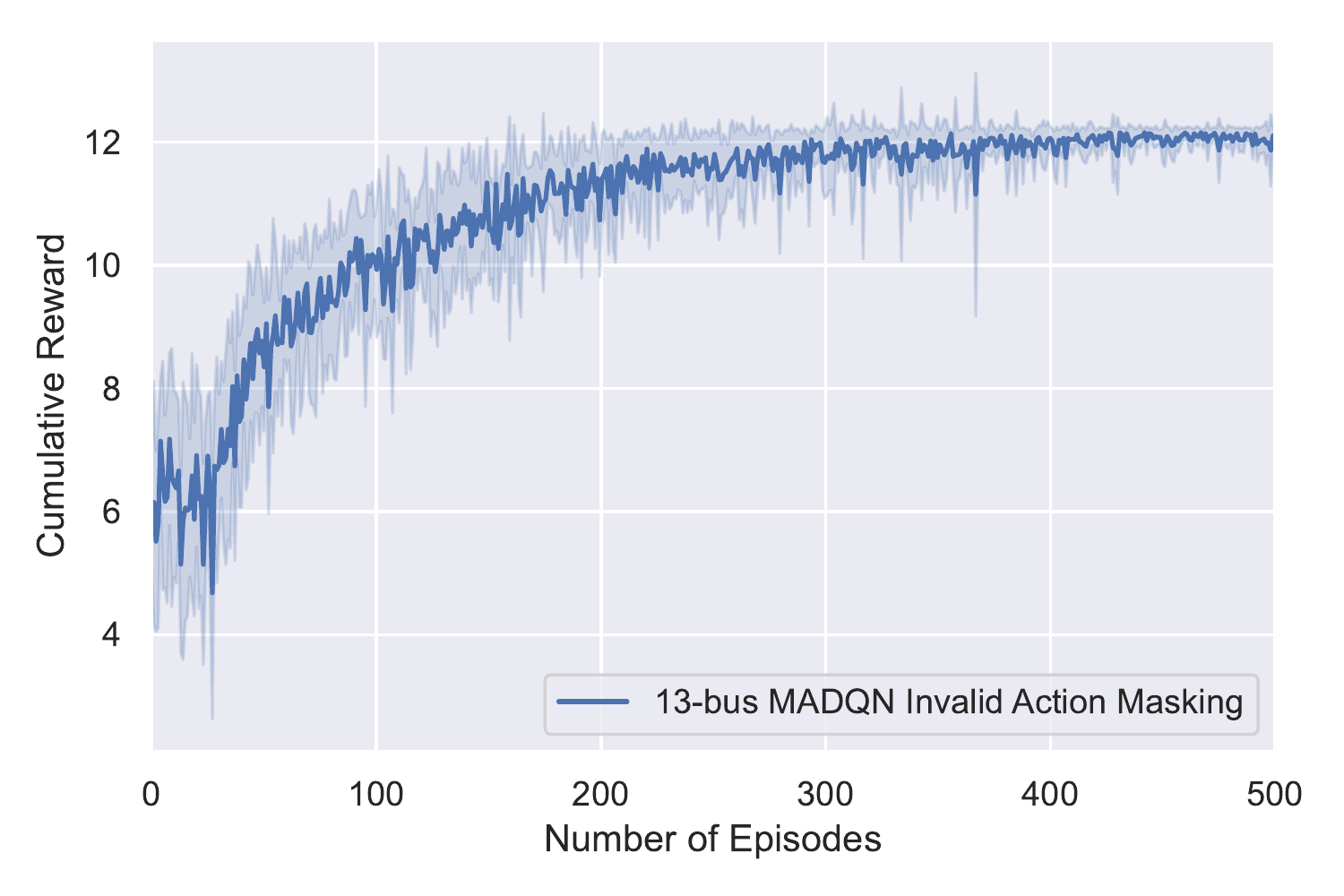}
	\caption{Training curve of multi-agent RL IEEE 13-node system.}
	\label{fig:301}
\end{figure}

The algorithm trains for 500 episodes and returns a cumulative reward curve shown in Fig. \ref{fig:301}. The cumulative reward represents the amount of reward that agents can achieve during each episode over the learning course. As can be seen, its value gradually increases until it reaches 12, which is the highest reward they could earn. After 300 episodes, the cumulative reward curve converges and the agents keep their same optimal policies.

\begin{figure}[!t]
	\centering
	\includegraphics[width=0.9\linewidth]{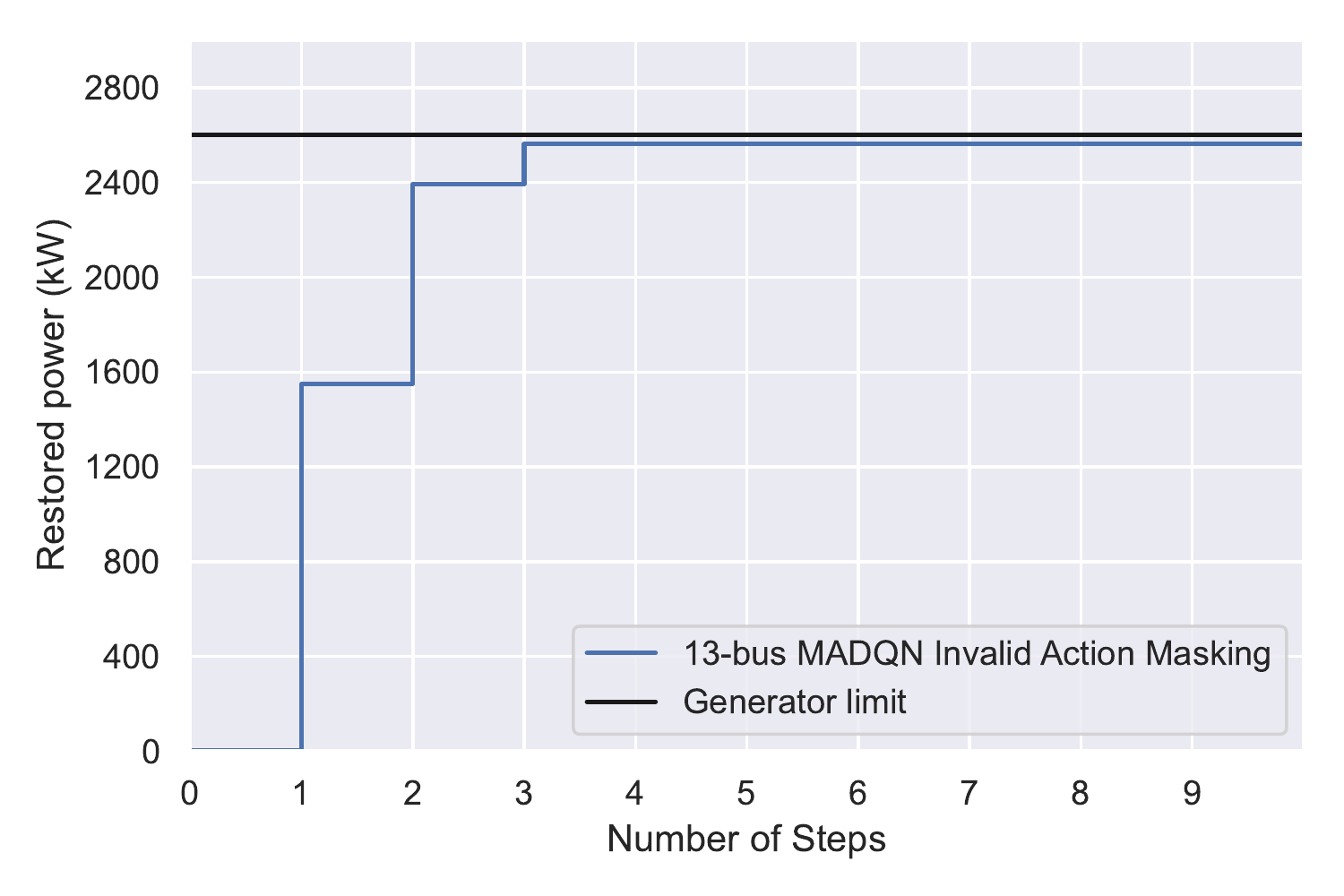}
	\caption{Testing of two-agent RL IEEE 13-node system.}
	\label{fig:302}
\end{figure}

Fig. \ref{fig:302} shows the testing result after training. The sequential configuration of circuit breakers is to restore loads 3, 7, 9, and 2 with the restored power of 2563 kW. As the maximum available generator is 2600 kW, the models, after being trained, can achieve up to 98.6\% of restored power after three steps.

\subsection{IEEE 123 Nodes}
\begin{figure}[!t]
	\centering
	\includegraphics[width=1.0\linewidth]{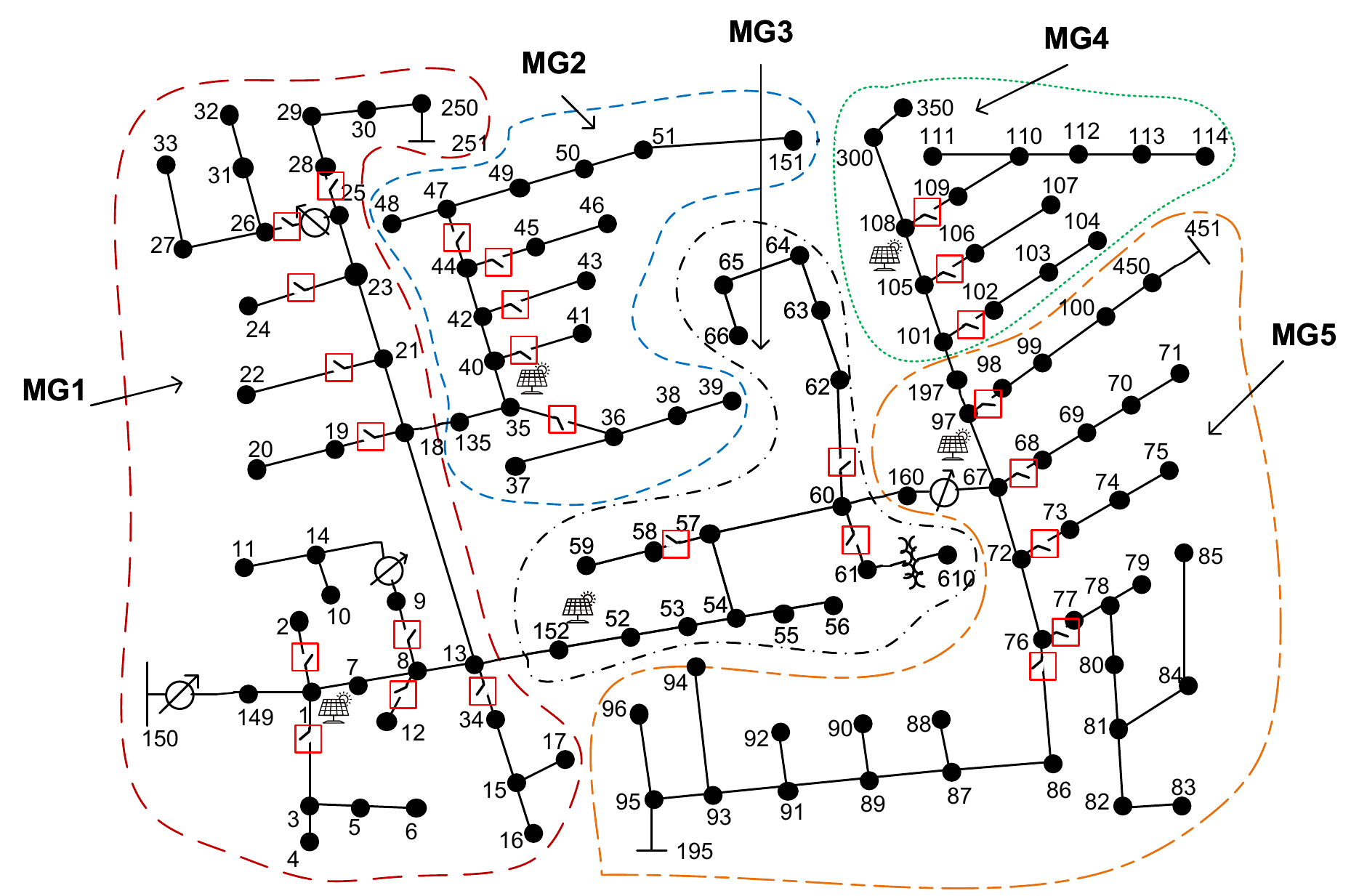}
	\caption{123-node networked microgrids configuration.}
	\label{fig:303}
\end{figure}

\noindent The algorithm is then expanded to validate the medium-size system of the IEEE 123-node distribution test feeder, as shown in Fig. \ref{fig:303}. In this case, the distribution system is divided into five microgrids corresponding to five agents. There are a total of 26 circuit breakers distributed throughout the system (10 circuit breakers in microgrid 1, 5 circuit breakers in microgrid 2, 3 circuit breakers in microgrid 3, 3 circuit breakers in microgrid 4, and 5 circuit breakers in microgrid 5)  to manage 3025 kW power of loads. Note that the number of circuit breakers in each microgrid is different depending on the number of loads and total power of each microgrid. A total of five renewable energy sources with a maximum capacity of 2400 kW are installed to provide power to the system during the outage.

\begin{figure}[!t]
	\centering
	\includegraphics[width=0.9\linewidth]{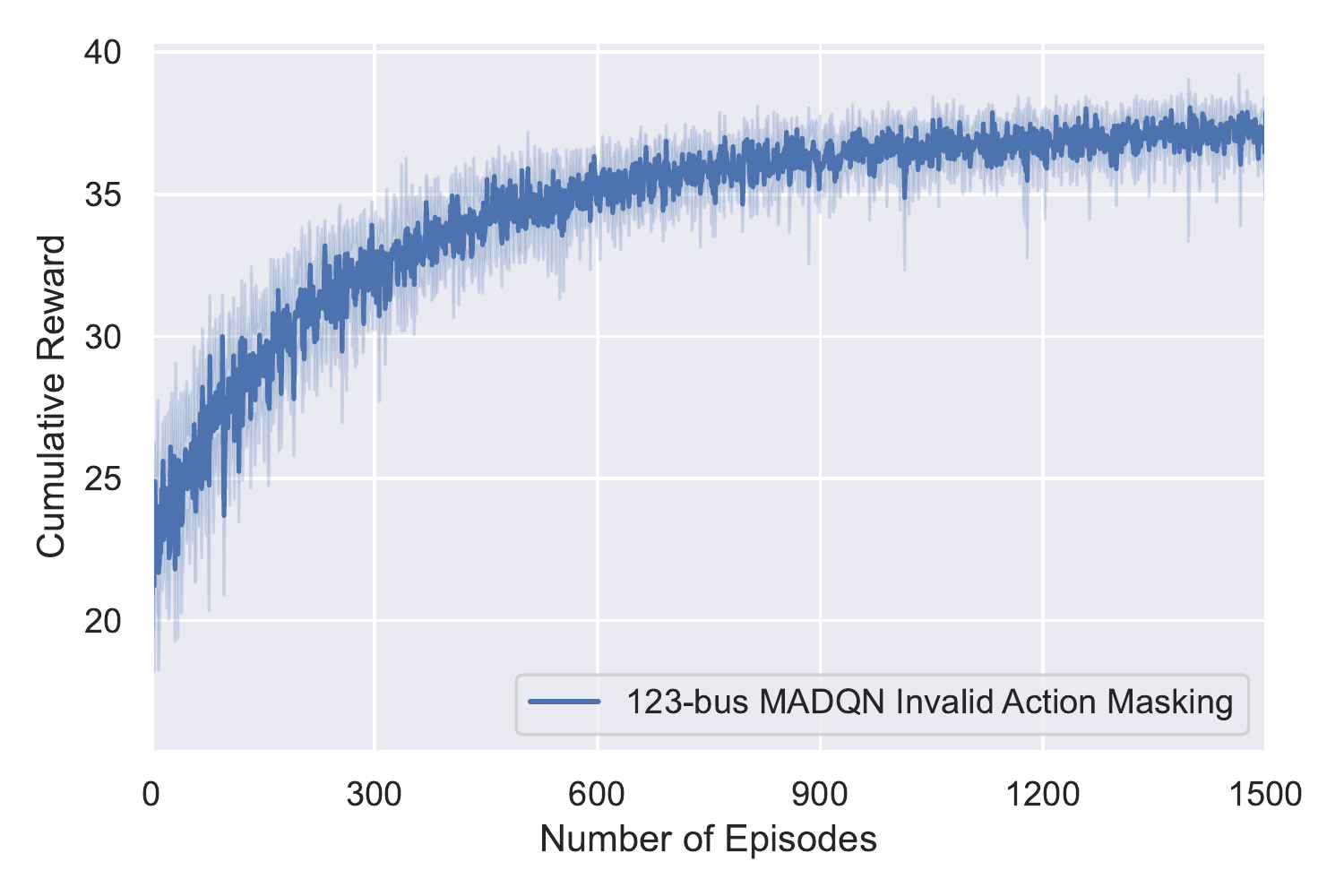}
	\caption{Training curve of multi-agent RL IEEE 123-node system.}
	\label{fig:304}
\end{figure}

The training curve of the IEEE 123-node system is shown in Fig. \ref{fig:304}. As can be seen, the cumulative reward converges (training curve) after 1500 episodes. Similar to the IEEE 13-node system's training curve, the cumulative reward increases as the agents explore and gain knowledge of the environment. The highest reward they can achieve in this setting is 37, with 7\% standard deviation. Using the trained models can help to recover up to 2305 kW of restored power, which is equivalent to 96.04\% of the 2400 kW total available power.

\subsection{IEEE 8500 Nodes}
\begin{figure}[!t]
	\centering
	\includegraphics[width=1.0\linewidth]{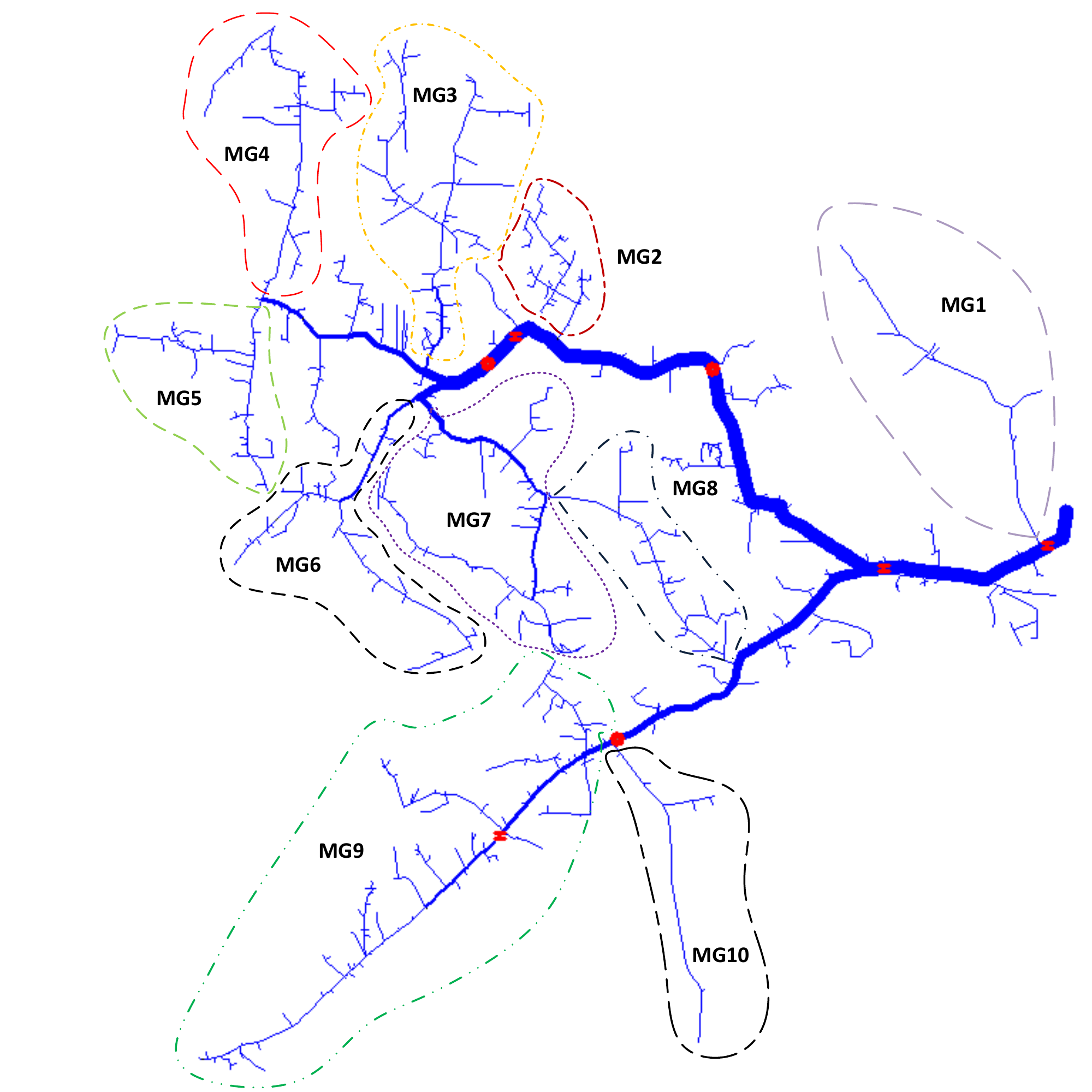}
	\caption{8500-node networked microgrids configuration.}
	\label{fig:305}
\end{figure}

\noindent The last case study is conducted with an IEEE 8500-node distribution test feeder, as shown in Fig. \ref{fig:305}. The distribution system is divided into ten microgrids with a total load of 4090 kW power. Multiple solar energy panels are located along the system with a total capacity of 2400 kW. The total number of circuit breakers being used in this system is 77, which is a very large number. In this case, we tried to use a single agent to make decisions; however, with the total action space being $2^{77}$, a single agent failed to address the problem.

\begin{figure}[!t]
	\centering
	\includegraphics[width=0.9\linewidth]{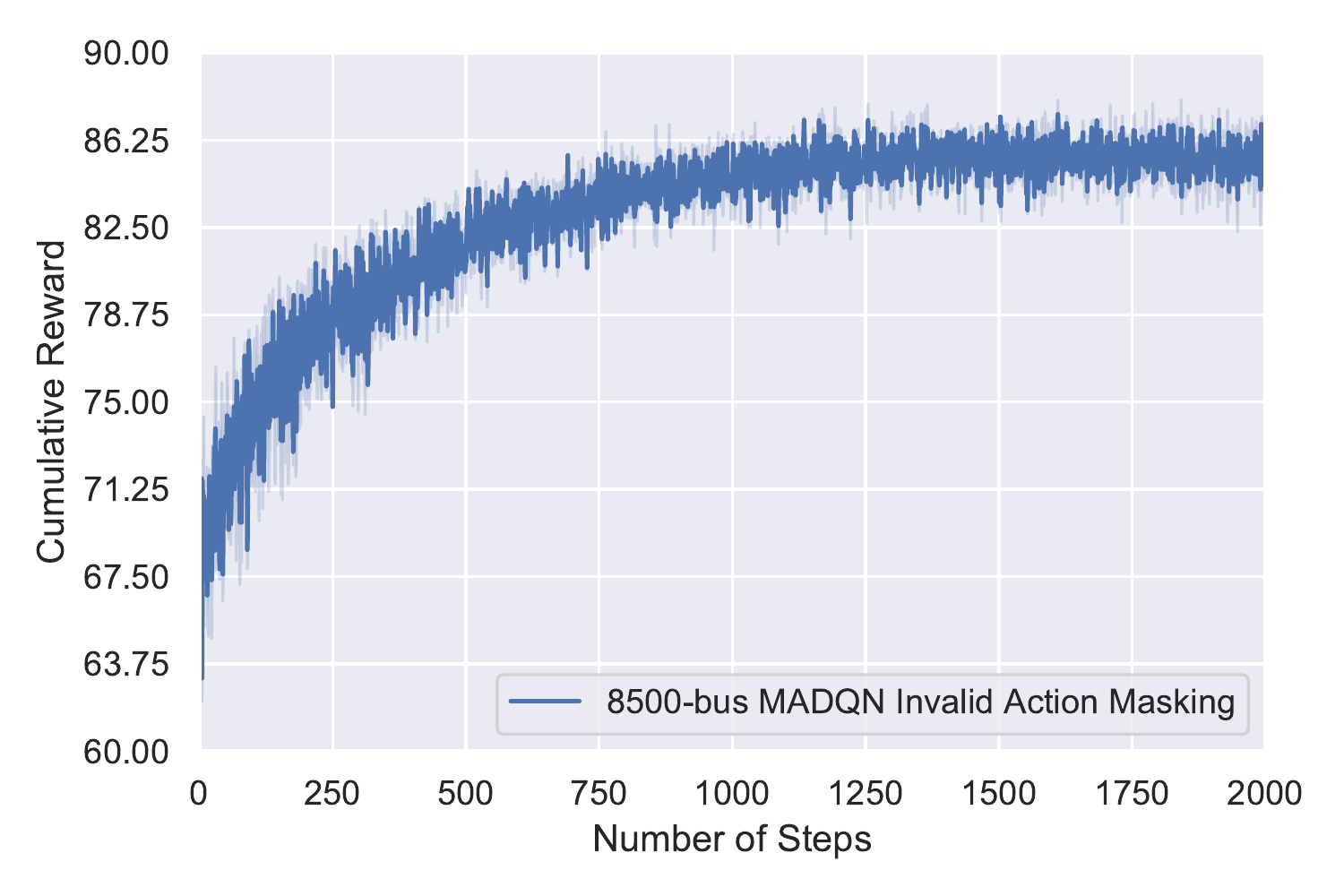}
	\caption{Training curve of multi-agent RL IEEE 8500-node system.}
	\label{fig:306}
\end{figure}

\begin{figure}[!t]
	\centering
	\includegraphics[width=0.9\linewidth]{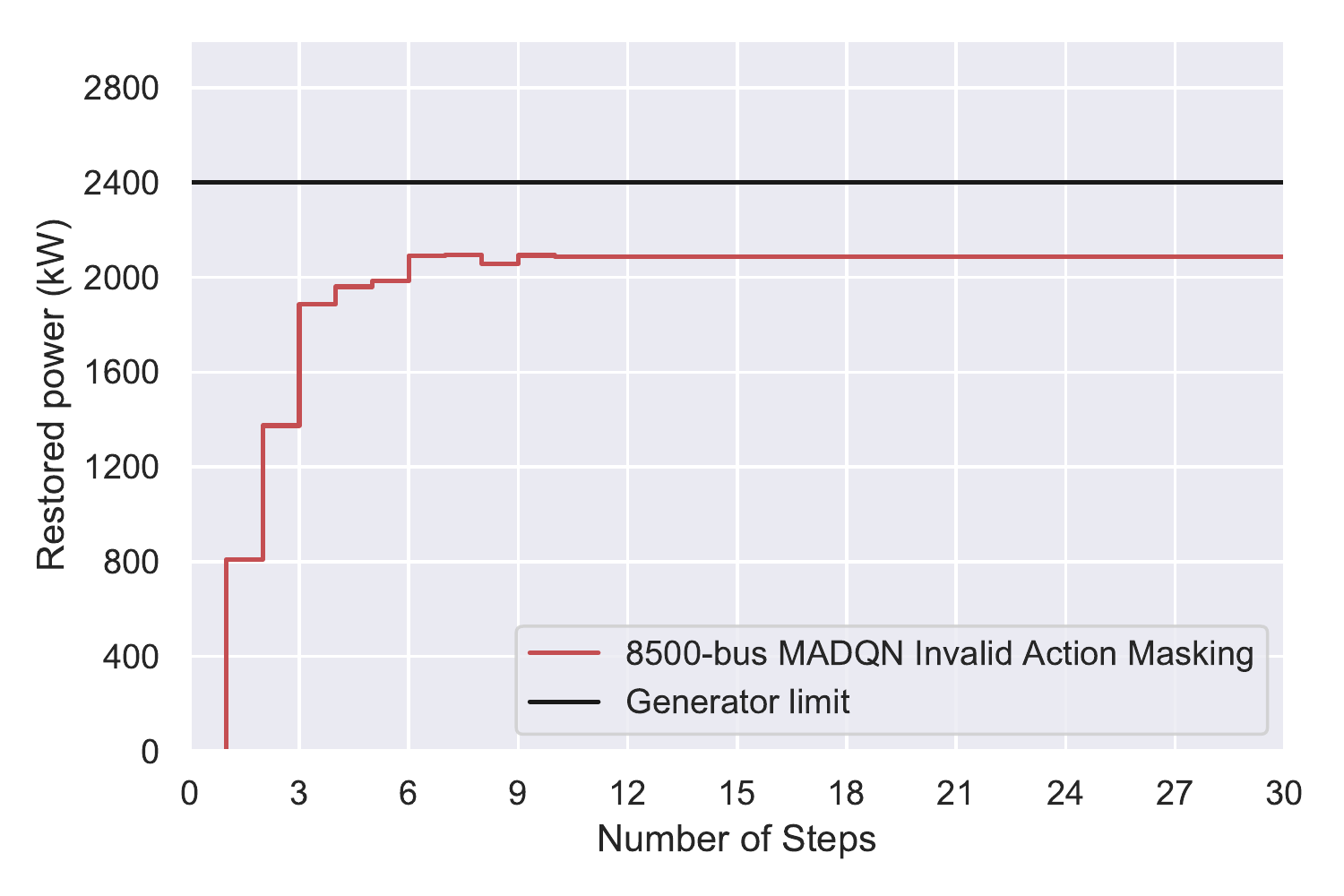}
	\caption{Testing of ten-agent RL IEEE 8500-node system.}
	\label{fig:307}
\end{figure}

Fig. \ref{fig:306} shows the cumulative reward result after training ten agents in 2000 episodes. As can be seen, the cumulative reward starts from 63.75 and gradually increases to 86.25, which is the optimal value of this setting. Fig. \ref{fig:307} shows the testing result for a total restoration of 2087.02 kW, which is equivalent to 86.96\% of the 2400 kW available power. Note that the available power from generators needs to supply for both restored loads and losses in the system.

\subsection{Comparative Analyses}

\subsubsection{Single-agent and multi-agent load restoration}
\noindent In this subsection, in order to fairly compare single-agent and multi-agent frameworks, all the common hyperparameters (ANN structure, mini-batch size, training episodes, etc.) are the same.

\begin{figure}[!t]
	\centering
	\includegraphics[width=0.9\linewidth]{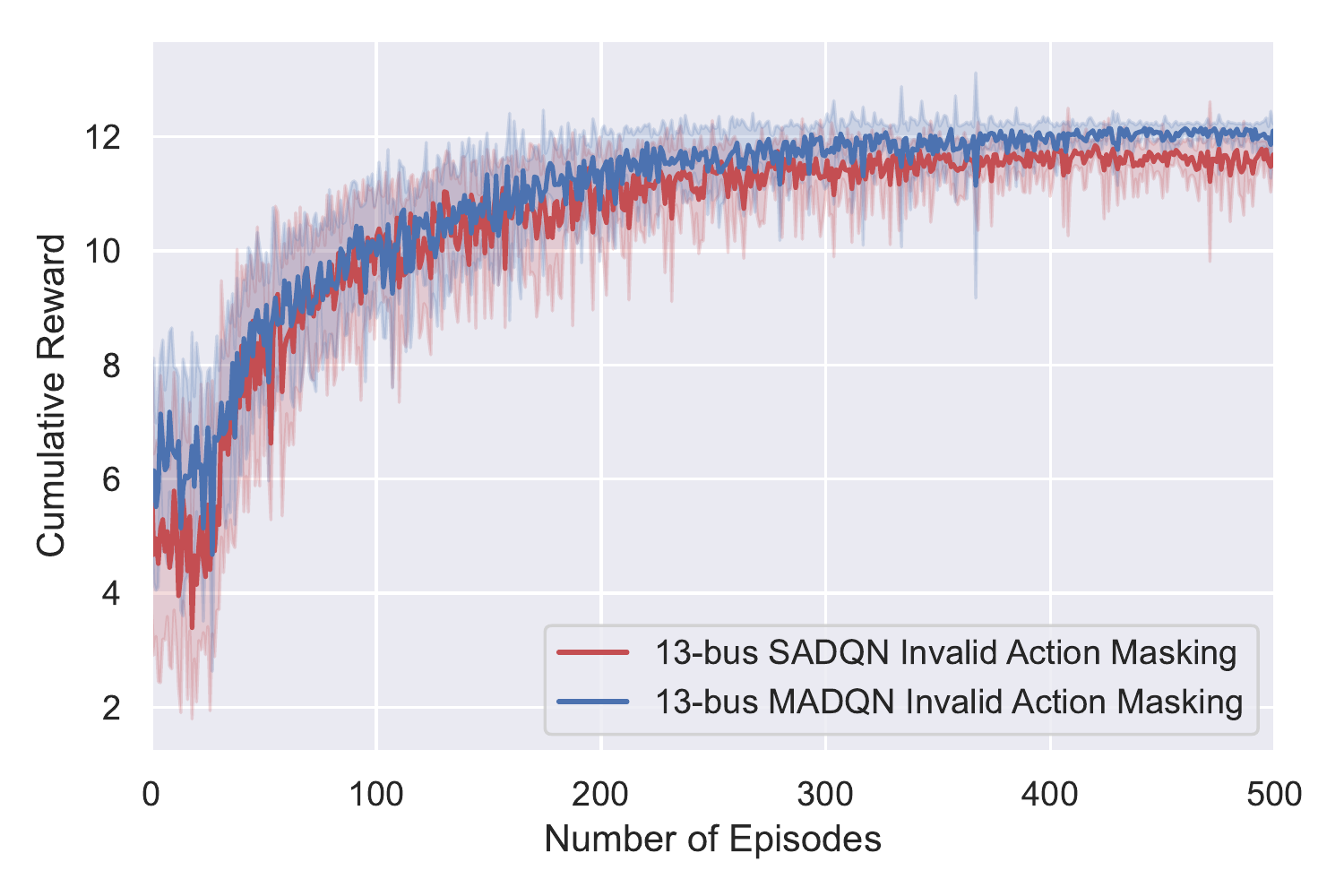}
	\caption{13-node single agent with and without action masking.}
	\label{fig:308}
\end{figure}

\begin{figure}[!t]
	\centering
	\includegraphics[width=0.9\linewidth]{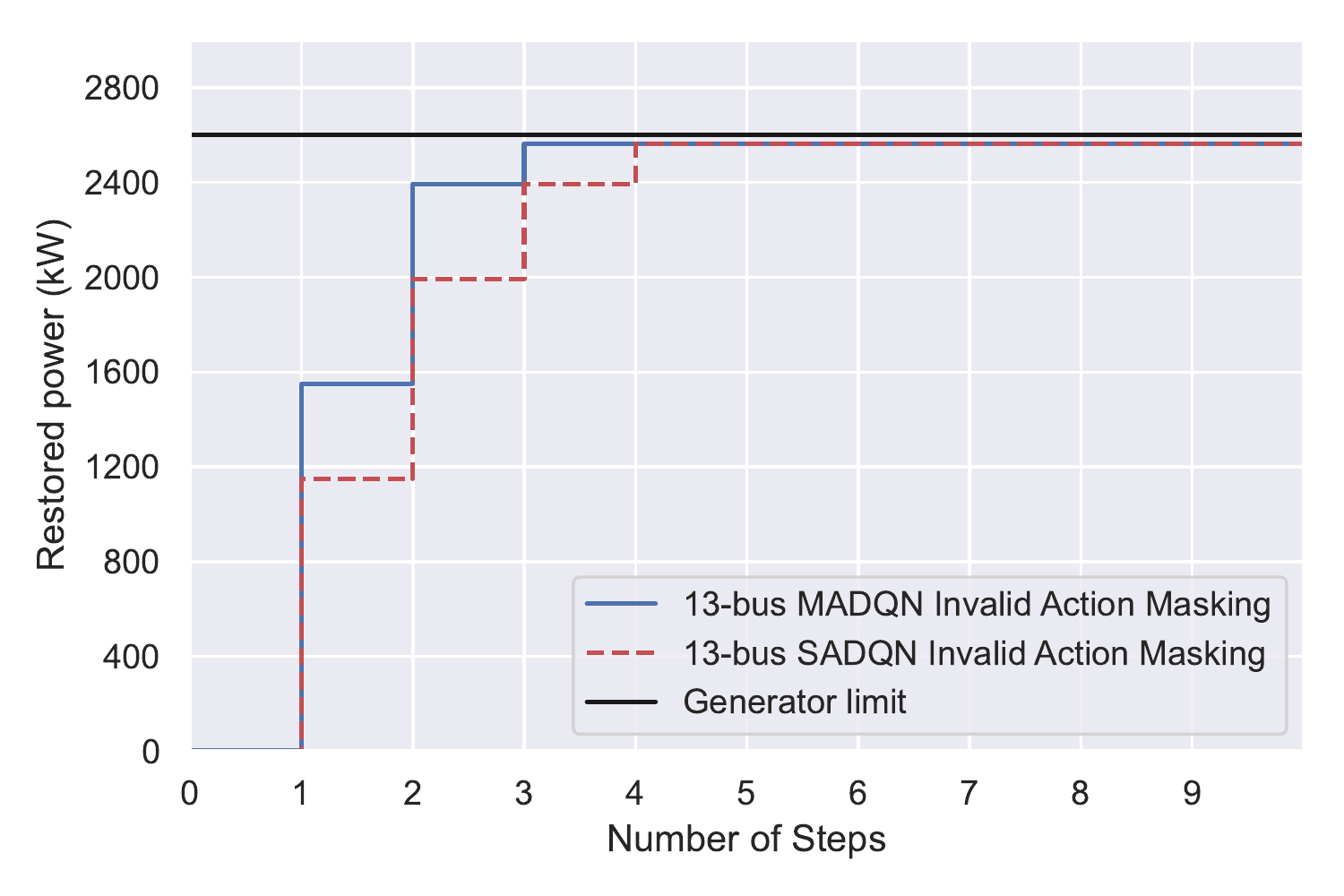}
	\caption{Testing results of two IEEE 13-node systems.}
	\label{fig:309}
\end{figure}

Fig. \ref{fig:308} shows the IEEE 13-node system results. For this small-size system, we can see that both frameworks can converge and find the optimal solution. Interestingly, the multi-agent can achieve better load pickup in each step, and obtain the optimal value one step earlier than a single agent as shown in Fig. \ref{fig:309}. The underlying reason is that the multi-agent framework enables better flexibility where each agent can control a small number of loads to be picked up in its area, and hence, it can quickly find the load to match the generation limit in its area. Whereas, the single agent has to control a much bigger number of loads, and thus, it needs more time to find the optimal solution.

\subsubsection{With and without invalid action masking technique}
\begin{figure}[!t]
	\centering
	\includegraphics[width=0.9\linewidth]{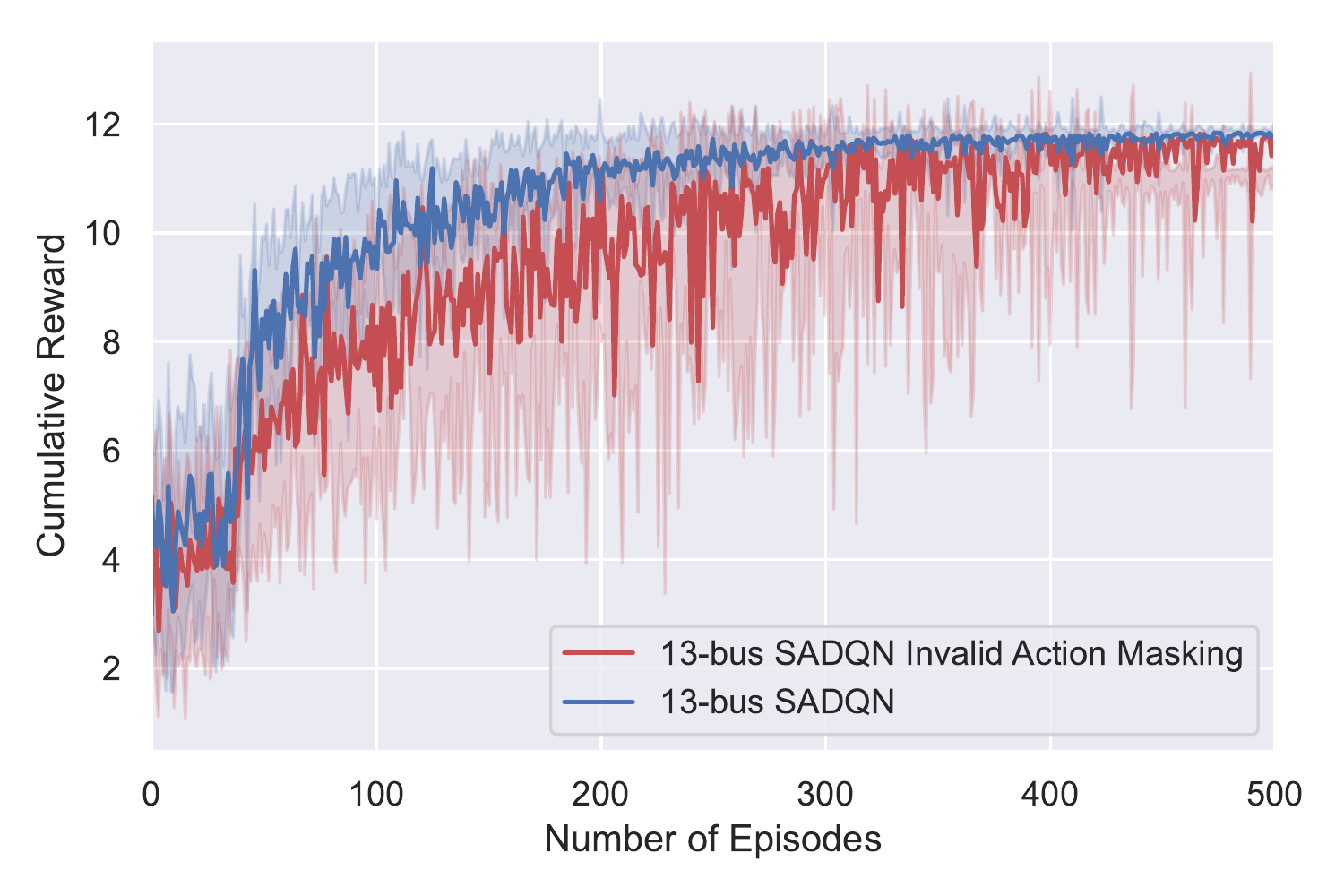}
	\caption{13-node single agent with and without invalid action masking.}
	\label{fig:310}
\end{figure}

\noindent The next comparison is carried out to analyze the effectiveness of invalid action masking in terms of dealing with safety constraints in reinforcement learning. In the first case, the IEEE 13-node system is used to apply for single agent with and without invalid action masking. As can be seen in Fig. \ref{fig:310}, the single agent with the invalid action masking technique can achieve the optimal reward faster (which can be clearly observed by the gap between the two curves) and has 9.75\% higher total cumulative reward than one without action masking (pure single agent). Note that the standard deviation of a single agent with invalid action masking is 57.32\% smaller than a pure single agent because the pure single agent fails to learn meaningful actions and easily drops the training reward value as it violates the power flow constraints.

\begin{figure}[!t]
	\centering
	\includegraphics[width=0.9\linewidth]{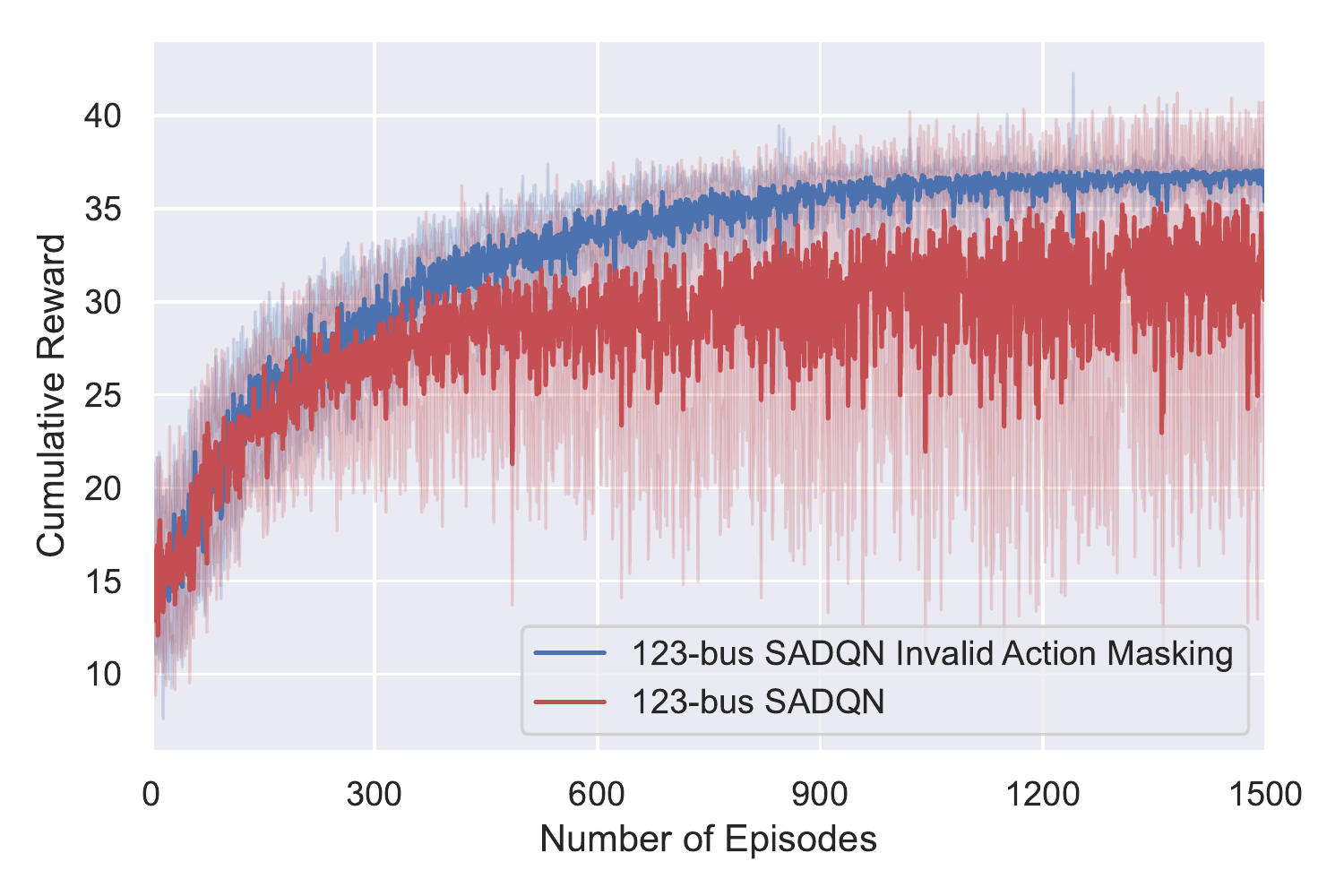}
	\caption{123-node single agent with and without invalid action masking.}
	\label{fig:311}
\end{figure}

In the second case, we expand the single-agent approach to an IEEE 123-node system. As can be seen from the training curve shown in Fig. \ref{fig:311}, although both approaches still can converge after 1500 episodes, result for the approach without invalid action masking can only reach a sub-optimal result (30) compared to the invalid action masking approach (37). In terms of standard deviation, the invalid action masking agent also has a lower value of 66\%, indicating that under the same settings, the invalid action masking technique definitely helps reduce exploration time and stabilize the learning process.

\begin{figure}[!t]
	\centering
	\includegraphics[width=0.9\linewidth]{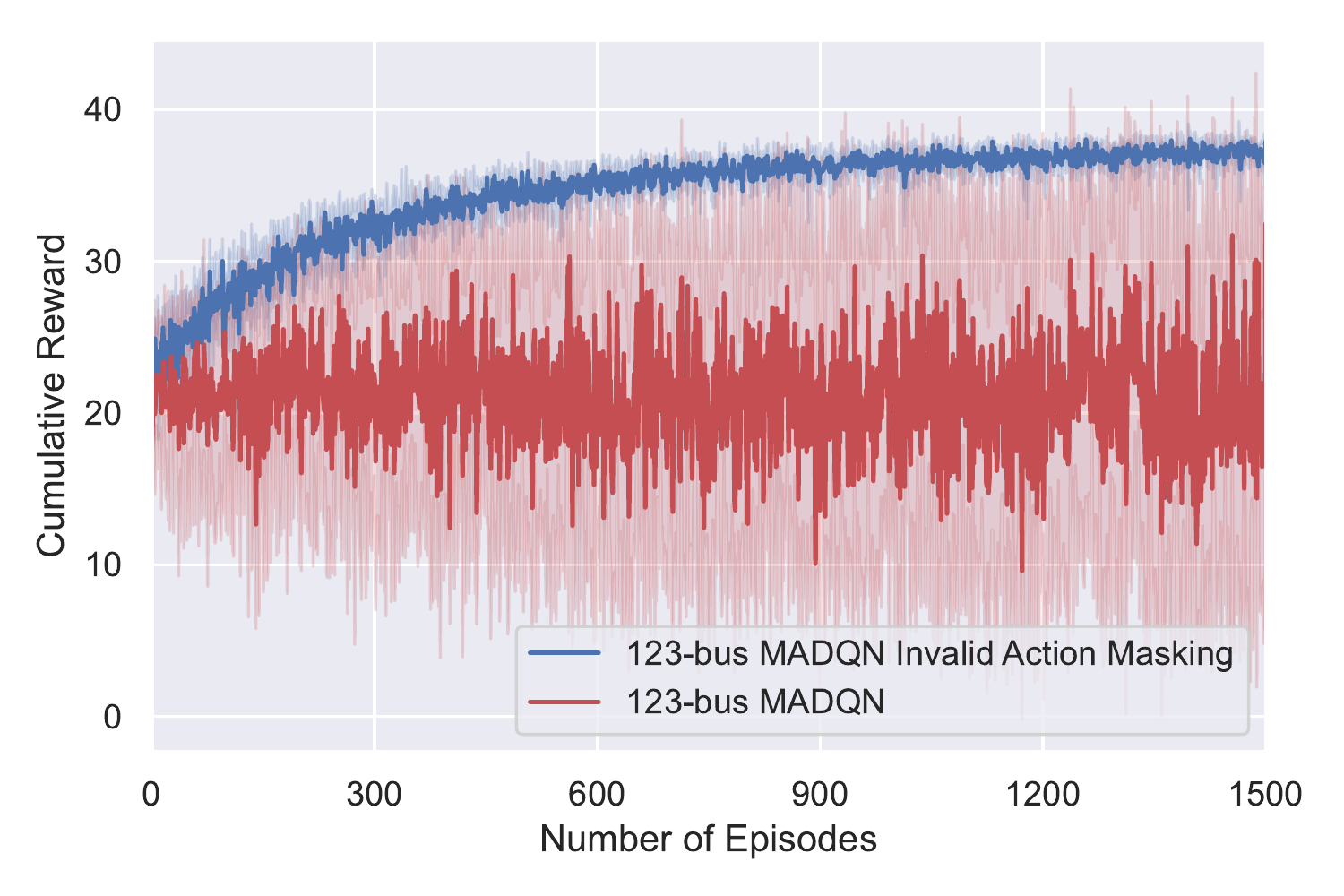}
	\caption{123-node multi agent with and without invalid action masking.}
	\label{fig:312}
\end{figure}

In the third scenario, we test the performance of the invalid action masking technique for a deep MARL framework using IEEE 123-node system. Fig. \ref{fig:312} depicts the critical importance of invalid action masking when the problem is scaled up. As can be seen, the learning curve of five pure agents starts at the value of 22 and doesn't increase during the training phase. This implies those agents cannot find any good policies given the fact that all other settings are the same compared to the one with invalid action masking. The high value of standard variance also dictates that the pure agents keep violating the constraints even at the end of the training phase.

\begin{figure}[!t]
	\centering
	\includegraphics[width=0.9\linewidth]{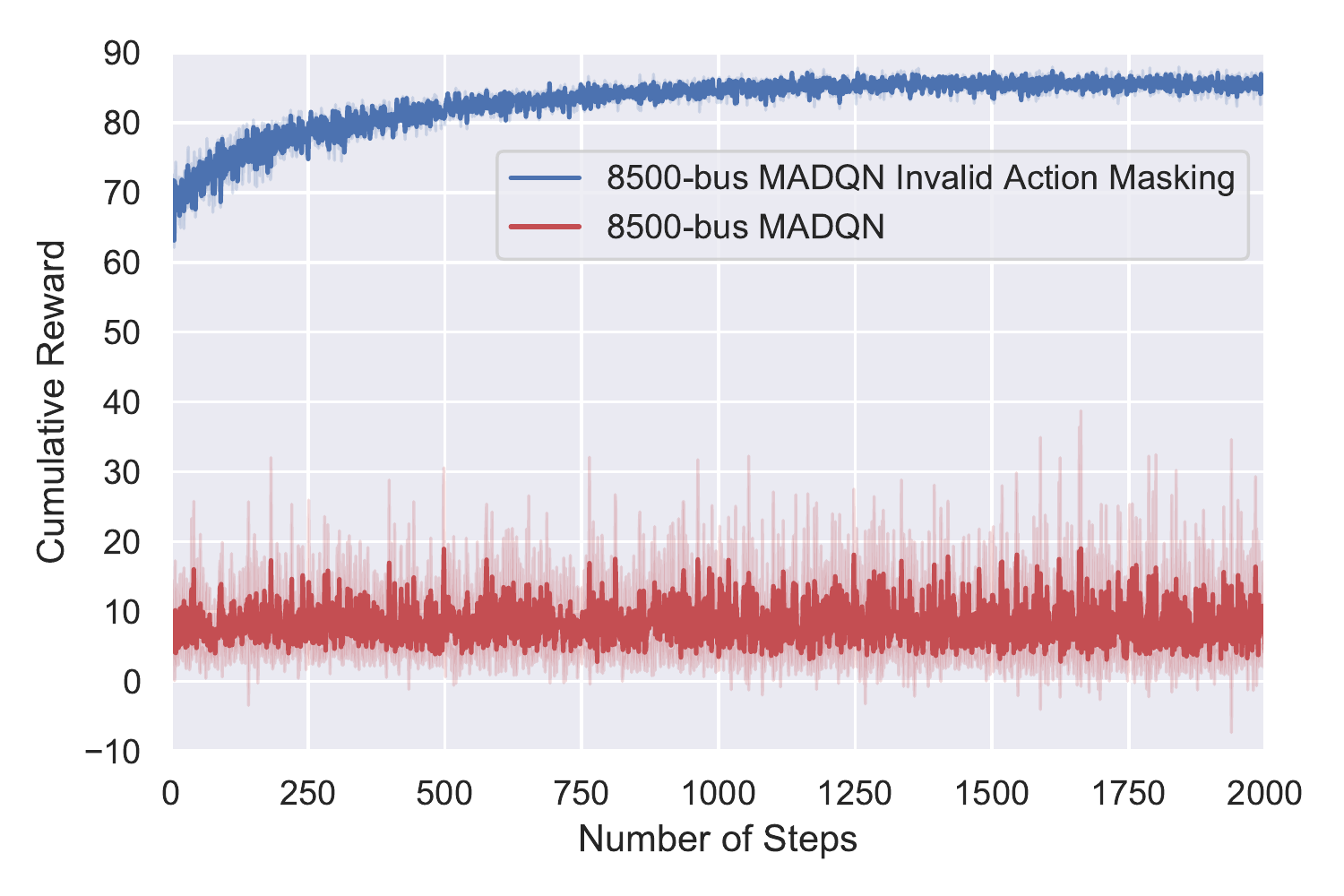}
	\caption{\textcolor{rev1}{8500-node multi agent with and without invalid action masking.}}
	\label{fig:313}
\end{figure}

\textcolor{rev1}{In the final scenario, we utilize the deep MARL framework that was previously applied to the IEEE 123-node system and expand it to the IEEE 8500-node system for comparison. As depicted in Fig. \ref{fig:313}, for the implementation of ten agents without invalid action masking, they faced difficulties collaborating and violated the physical constraints. This led to an unconverged policy, unlike the scenario where invalid action masking was used with the same settings.}

\textcolor{rev1}{Table \ref{tab:mytab1} provides a comprehensive comparison of single-agent and multi-agent approaches in terms of solution time and restored load under different system sizes. The results demonstrate the unique effectiveness of our proposed approach in improving the stability of learning convergence and the cumulative reward in large-scale settings.}

\begin{table}[!t]
	\textcolor{rev1}{
		\centering
		\caption{Comparison of Solution Time and Restored Load With and Without Invalid Action Masking. NC: Not converged}
		\label{tab:mytab1}
		\resizebox{\columnwidth}{!}{%
			\begin{tabular}{cccccc}
				\cline{3-6}
				\multicolumn{2}{c}{\multirow{2}{*}{}}                                      & \multicolumn{2}{c}{\begin{tabular}[c]{@{}c@{}}Solution\\ Time (s)\end{tabular}} & \multicolumn{2}{c}{\begin{tabular}[c]{@{}c@{}}Restored \\ Load (\%)\end{tabular}}                                                                                                                          \\ \cline{3-6}
				\multicolumn{2}{c}{}                                                       & \begin{tabular}[c]{@{}c@{}}Single\\ Agent\end{tabular}                          & \begin{tabular}[c]{@{}c@{}}Multi\\ Agent\end{tabular}                             & \begin{tabular}[c]{@{}c@{}}Single\\ Agent\end{tabular} & \begin{tabular}[c]{@{}c@{}}Multi\\ Agent\end{tabular}         \\ \hline
				\multirow{2}{*}{\begin{tabular}[c]{@{}c@{}}IEEE \\ 13-node\end{tabular}}   & Without Mask                                                                    & 0.43                                                                              & 0.35                                                   & 98.58                                                 & 98.58 \\
				                                                                           & With Mask                                                                       & 0.41                                                                              & 0.30                                                   & 98.58                                                 & 98.58 \\ \hline
				\multirow{2}{*}{\begin{tabular}[c]{@{}c@{}}IEEE \\ 123-node\end{tabular}}  & Without Mask                                                                    & 2.43                                                                              & NC                                                     & 91.88                                                 & NC    \\
				                                                                           & With Mask                                                                       & 2.17                                                                              & 1.00                                                   & 96.04                                                 & 96.04 \\ \hline
				\multirow{2}{*}{\begin{tabular}[c]{@{}c@{}}IEEE \\ 8500-node\end{tabular}} & Without Mask                                                                    & NC                                                                                & NC                                                     & NC                                                    & NC    \\
				                                                                           & With Mask                                                                       & 12.27                                                                             & 2.26                                                   & 64.71                                                 & 86.96 \\ \hline
			\end{tabular}%
		}}
\end{table}

\subsubsection{Limitations}
\noindent Although our experimental results have demonstrated promising performance, there are some limitations that need to be further investigated. (1) Since the input and output of the model after training are fixed, the model cannot be reused in case additional circuit breakers are installed that modify the structure of the distribution system. (2) Although the algorithm can execute the testing configuration of circuit breakers almost immediately, the training process may take a long time because agents need sufficient time to explore the environment, and the training time increases as the number of agents grows. This training time drawback may be reduced by leveraging the parallel training feature that most reinforcement learning platforms provide \cite{pmlr-v80-liang18b, terry2020pettingzoo, https://doi.org/10.48550/arxiv.2206.10558}. Since reinforcement learning emerged only recently for addressing problems in power systems, those limitations can become subjects for further study.

\section{Conclusion}
\noindent In this work, we proposed a multi-agent deep reinforcement learning approach to improve the resilience of distribution systems after a power outage. The load restoration agents were trained offline in a centralized way to have better information. Then the trained agents were used at execution time in a decentralized manner. The invalid action masking technique was used to guarantee the power flow constraints both in the training and testing phases. The experimental results show that agents can collaborate to achieve the highest reward and the learning curve is stable during the training phase. Two major disadvantages were pointed out, namely generalization and adaptability; those issues will be our follow-up research.


\ifCLASSOPTIONcaptionsoff
	\newpage
\fi

\bibliographystyle{IEEEtran}
\bibliography{Paper}

\newpage 
\end{document}